\documentstyle{article}
\newcount\Mac  \Mac=0  
\renewcommand{\thefootnote}{\fnsymbol{footnote}}
\newcommand{\PhiB}{\Phi_{^8\!\rm B}}
\newcommand{\PhiBe}{\Phi_{^7\!\rm Be}}
\newcommand{\PhiCNO}{\Phi_{\rm CNO}}
\newcommand{\Phip}{\Phi_{\rm p}}
\newcommand{\degree}{^{{\circ}}}
\renewcommand{\theequation}{\thesection.\arabic{equation}}
\newcommand{\reseteq}{\setcounter{equation}{0}}
\newcommand{\mnuLR}{m_{LR}}
\newcommand{\mnuLL}{m_{LL}}
\newcommand{\mnuRR}{m_{RR}}

\def\noCl{{\begin{picture}(0.4,0)(0,0)\put(0,0){\rm Cl}\put(0,-0.05){\line(1,1){0.4}}\end{picture}}}
\def\noGa{{\begin{picture}(0.4,0)(0,0)\put(0,0){\rm Ga}\put(0,-0.05){\line(1,1){0.4}}\end{picture}}}
\def\noSK{{\begin{picture}(0.4,0)(0,0)\put(0,0){\rm SK}\put(0.03,-0.05){\line(1,1){0.4}}\end{picture}}}

\def\Red{}
\def\Black{}
\def\Blue{}
\def\Green{}
\newcommand{\lascia}[1]{}
\def\puttag(#1,#2)#3{\put(#1,#2){\makebox(0,0){\rm\Blue #3\Black}}}
\newcommand{\tag}[1]{{\Blue\rm #1\Black}}

\def\circa#1{\,\raise.3ex\hbox{$#1$\kern-.75em\lower1ex\hbox{$\sim$}}\,}

\newcommand{\riga}[1]{\noalign{\hbox{\parbox{\textwidth}{#1}}}\nonumber}
\newcommand{\psfigure}[1]{\ifnum\Mac=1 \special{picture #1} \else \includegraphics{#1} \fi}
\def\putps(#1,#2)(#3,#4)#5{\ifnum\Mac=1 \put(#1,#2){\special{picture #5}}
\else  \put(#3,#4){\includegraphics{#5}} \fi}

\def\One{\hbox{1\kern-.24em I}}

\newcommand{\eV}{\,{\rm eV}}
\newcommand{\cm}{\,{\rm cm}}
\newcommand\Ord{{\cal O}}
\newcommand{\eps}{\varepsilon}

\newcommand{\eq}[1]{~{\rm (\ref{eq:#1})}}

\newcommand{\NP}{Nucl. Phys.}
\newcommand{\PRL}{Phys. Rev. Lett.}
\newcommand{\PL}{Phys. Lett.}
\newcommand{\PR}{Phys. Rev.}
\makeatletter
\def\art{\@ifnextchar[{\eart}{\oart}}
\def\eart[#1]#2#3#4#5#6{{\rm #2}, {\em #3 \bf #4} {\rm (#6) #5} ({\em #1})}
\def\hepart[#1]#2{{\rm #2, \em#1}}
\newcommand{\oart}[5]{{\rm #1}, {\em #2 \bf #3} {\rm (#5) #4}}
\newcounter{alphaequation}[equation]
\def\thealphaequation{\theequation\hbox to
0.6em{\hfil\alph{alphaequation}\hfil}}
\def\eqnsystem#1{
\def\@eqnnum{{\rm (\thealphaequation)}}
\def\@@eqncr{\let\@tempa\relax \ifcase\@eqcnt \def\@tempa{& & &} \or
  \def\@tempa{& &}\or \def\@tempa{&}\fi\@tempa
  \if@eqnsw\@eqnnum\refstepcounter{alphaequation}\fi
\global\@eqnswtrue\global\@eqcnt=0\cr}
\refstepcounter{equation} \let\@currentlabel\theequation \def\@tempb{#1}
\ifx\@tempb\empty\else\label{#1}\fi
\refstepcounter{alphaequation}
\let\@currentlabel\thealphaequation
\global\@eqnswtrue\global\@eqcnt=0 \tabskip\@centering\let\\=\@eqncr
$$\halign to \displaywidth\bgroup \@eqnsel\hskip\@centering
$\displaystyle\tabskip\z@{##}$&\global\@eqcnt\@ne
\hskip2\arraycolsep\hfil${##}$\hfil& \global\@eqcnt\tw@\hskip2\arraycolsep
$\displaystyle\tabskip\z@{##}$\hfil
\tabskip\@centering&\llap{##}\tabskip\z@\cr}

\def\endeqnsystem{\@@eqncr\egroup$$\global\@ignoretrue} \makeatother

\begin{document}
\begin{quote}
{\em June 1998}\hfill {\bf UCB-PTH-98/44}\\
hep-ph/9807235 \hfill{\bf LBNL-42020}\\
{\bf IFUP-TH/25-98 \hfill SNS-PH/98-15}\\  
\end{quote}
\bigskip
\centerline{\huge\bf\Red Oscillations of solar and atmospheric neutrinos%
\footnote[2]{This work was supported
in part by the U.S. Department of Energy under Contracts DE-AC03-76SF00098,
in part by the National Science Foundation under grant PHY-95-14797,
in part by the TMR network under the EEC contract n.\ ERBFMRX-CT960090.}}
\bigskip\bigskip\Black
\begin{center}\large
{\bf R.\ Barbieri$^1$,  L.J.\ Hall$^2$, D.\ Smith$^2$, A.\ Strumia$^3$ {\rm and} N.\ Weiner$^2$}\\
\bigskip{\em
$^1$ Scuola Normale Superiore, and\\
     INFN, sezione di Pisa,  I-56126 Pisa, Italia\\[3mm]

$^2$ Department of Physics and\\
     Theoretical Physics Group, Lawrence Berkeley National Laboratory\\
     University of California, Berkeley, California 94720\\[3mm]

$^3$ Dipartimento di fisica, Universit\`a di Pisa, and\\
     INFN, sezione di Pisa,  I-56127 Pisa, Italia}\\[8mm]

\bigskip\bigskip\Blue
\end{center}
\centerline{\large\bf Abstract}
\begin{quote}\large\indent
Motivated by recent results from SuperKamiokande, we study both solar 
and atmospheric neutrino
fluxes in the context of oscillations of the three known neutrinos.
We aim at a global view which identifies the various possibilities,
rather than attempting the most accurate determination of the
parameters of each scenario.
For solar neutrinos we emphasise the importance of performing a general analysis, 
independent of any particular
solar model and we consider the possibility that any one of the
techniques ---  chlorine, gallium or water Cerenkov ---  has a large unknown
systematic error, so that its results should be discarded. The atmospheric neutrino 
anomaly is studied by paying special attention to the ratios of upward and
downward going $\nu_e$ and $\nu_\mu$ fluxes. Both anomalies can be 
described in a minimal scheme where the respective oscillation frequencies are
widely separated or in non-minimal schemes with two comparable oscillation frequencies.

We discuss explicit forms of
neutrino mass matrices in which both atmospheric and solar neutrino fluxes
are explained.
In the minimal scheme we identify
only two `zeroth order' textures that can result from unbroken symmetries.
Finally we discuss experimental strategies for the determination of the 
various oscillation parameters.
\end{quote}\Black
\thispagestyle{empty}\newpage
\renewcommand{\thefootnote}{\arabic{footnote}}
\setcounter{page}{1}

\section{Introduction}
The solar and atmospheric neutrino flux anomalies have both been considerably strengthened by recent
observations from Super-Kamiokande.  The solar neutrino flux is measured to be~\cite{SK-solar}
$0.37 \pm 0.03$ of that
expected from the `BP95' standard solar model~\cite{BP},
without including any theoretical error.
This is the fifth solar neutrino experiment to
report results in strong disagreement with the predictions of solar models. Furthermore, using a solar
model independent analysis, the measured solar fluxes are found to be in conflict with each other. For
events at SuperKamiokande with visible energies of order a GeV, the ratio of 1 ring $\mu$-like to
$e$-like events is $0.66 \pm 0.10$ that expected from calculations of the flux of neutrinos produced in
the atmosphere in cosmic ray showers~\cite{SK-atm}. Furthermore, the distribution in zenith angle of
these 1 ring events provides striking evidence for a depletion of $\nu_\mu$ which depends on the
distance travelled by the neutrinos before reaching the Super-Kamiokande detector. In particular, the
observed up/down ratio of the multi-GeV, $\mu$-like events is $0.52 \pm 0.07$.   This significantly
strengthens the evidence that $\nu_\mu$ oscillate as they traverse the earth.

In this paper, we interpret the solar and atmospheric neutrino flux anomalies in terms of oscillations
of the three known neutrinos
$\nu_{e, \mu, \tau}$. The lightness of these three neutrinos, relative to the charged fermions, can be
simply understood as resulting from large ${\rm SU}(2)_L \otimes {\rm U}(1)_Y$ invariant masses for the right-handed
neutrinos, via the see-saw mechanism. We do not consider the possibility of a fourth light neutrino, as
it would have to be singlet under ${\rm SU}(2)_L \otimes {\rm U}(1)_Y$, and would either require a new mass scale far
below the weak scale, running counter to the idea of the see-saw mechanism, or a more complicated
see-saw.

Theoretical ideas about generation mixing are guided by the quark sector, where the mixing angles are
all small, indicating a hierarchical breaking of horizontal symmetries in nature. A similar hierarchy
of horizontal symmetry breaking in the lepton sector is also likely to yield small angles, suggesting
small probabilities for a neutrino to oscillate from one flavour to another. However, the solar and
atmospheric neutrino flux measurements both require neutrino survival probabilities, $P_{ee}$ and
$P_{\mu \mu}$, far from unity. Over a decade ago~\cite{MSW}, it was realised that large angles were not necessary
to account for the large suppression of solar neutrino fluxes --- while $\nu_e$ have charged current
interactions in the solar medium, $\nu_{\mu, \tau}$ do not, allowing a level crossing phenomena where a
$\nu_e$ state produced in the solar interior evolves to a $\nu_{\mu,\tau}$ state as it traverses the
sun. This simple picture can reconcile the three types of solar neutrino flux measurements with the
standard solar model, for a mixing angle as small as 0.03 --- a significant achievement. Could such
resonant oscillations occur for atmospheric neutrinos in the earth, again allowing a small vacuum
mixing angle? In this case, since the earth does not have a continuously varying  density,
the matter mixing angle in the earth is much larger than the vacuum mixing
angle only in a small range of energies. 
Hence, an oscillation interpretation of
the atmospheric neutrino fluxes requires a large mixing angle, and calls into question the frequently
stated theoretical prejudice in favour of small mixing angles.

\medskip

In this paper, we attempt to understand both solar and atmospheric neutrino fluxes using
3-generation neutrino oscillations,
aiming at a global view which identifies the various possibilities,
rather than attempting the most accurate determination of the
parameters of each scenario.
When data from chlorine, gallium and water Cerenkov detectors
are fitted to a standard solar model, standard analyses find very small regions of neutrino mass
and mixing parameters. For 2-generation mixing, these are known as the ``small angle MSW'', ``large
angle MSW'' and ``just so'' regions. This analysis has been extended to the case of three
generations~\cite{sun3gen}, with a single matter resonance in the sun, as suggested by the atmospheric
neutrino data. The large and small angle MSW areas are found to merge into a single MSW volume of
parameter space.
In subsection~\ref{SSMindep}, we study how this volume is enlarged when a solar model independent
analysis of the solar fluxes replaces the use of a single solar model.
In subsection~\ref{OneExpWrong} we extend our
analysis to see what areas of neutrino parameter space become allowed if one of the three
observational techniques to measure the solar fluxes is seriously in error.

We combine these regions of parameters with those yielding the atmospheric fluxes, and find there is
still considerable allowed ranges of masses and mixing angles.
This is done in section~\ref{AtmStandard},
assuming that the smallest of the two neutrino squared mass differences is
too small to affect the oscillations of atmospheric neutrinos (minimal scheme).
In section~\ref{AtmNonStandard}, on the contrary, we allow for the possibility that the two independent
neutrino squared mass differences are both large enough
to affect atmospheric neutrino oscillations (non minimal schemes).
For solar neutrinos, this requires that there is a serious flaw either in at least one measurement
technique or in solar model analyses.

The forms of neutrino mass matrices that can lead to a large $\nu_\mu \rightleftharpoons
\nu_\tau$ mixing for atmospheric neutrinos are discussed in section~\ref{matrices}.
In section~\ref{models}
only two `zeroth order' textures for neutrinos masses are identified 
that can account for the atmospheric and solar neutrino data in the
minimal scheme and can result from unbroken symmetries.

Our conclusions are drawn in section~\ref{Conclusions}.
Based on a simple set of alternative hypotheses,
we discuss how future measurements could eventually determine 
the two neutrino mass differences and 
the three mixing angles.

\section{Solar neutrinos: model-independent analysis}\reseteq\label{Sun}
In the flavour eigenstate basis, in which
the charged lepton mass matrix is  diagonal, the neutrino mass matrix is in general non-diagonal. It
may be  diagonalized by a unitary transformation:
\begin{equation}
\nu_f = V^*_{fi} \nu_i
\label{eq:Vfields}
\end{equation} where $\nu_f$ and $\nu_i$ are flavour and mass eigenstate fields, respectively. The
leptonic analogue of the Cabibbo-Kobayashi-Maskawa mixing matrix is 
$V^T$, since the $W$ boson couples to the charged current $\bar{\nu}_{i_L} V^T_{if} \gamma^\mu
e_{f_L}$. In addition to the three Euler angles, $V$  contains physical phases: one if the light
neutrinos are Dirac, and three  if they are Majorana. These flavour and mass eigenstate fields destroy
basis  states which are related by
\begin{equation} |\nu_f\rangle = V_{fi} |\nu_i\rangle
\label{eq:Vstates}
\end{equation}
If some process creates a flavour eigenstate, $|\nu_f\rangle$, at time $t=0$,  then at a
later time $t$ it will have evolved to the state
$|\nu_f,t\rangle = \psi_{f'}(t) |\nu_{f'}\rangle$ via the matrix Schroedinger equation
\begin{equation} i { d \psi \over dt} = (V \frac{m_\nu^2}{2E} V^\dagger + A_{\rm CC}+{\cal E}) \psi
\label{eq:SE}
\end{equation}
where $E$ is the energy of the relativistic neutrino,
$m_\nu$ is the diagonal neutrino mass matrix with entries $m_i$,
${\cal E}$ is an irrelevant term proportional to the unit matrix,
and $A_{\rm CC}$ represents matter effects.
For neutrinos propagating in matter with electron number density $N_e$,
$A_{\rm CC}$ is a matrix with a single non-zero entry,
$A_{\rm CC}^{11} =\sqrt{2} G_F N_e$.

The mixing matrix $V$ can be written quite generally as
\begin{equation} V =
R_{23}(\theta_{23}) \pmatrix{1 &0&0 \cr 0&e^{i \phi}&0 \cr 0&0&1} 
R_{13}(\theta_{13})
R_{12}(\theta_{12})
\pmatrix{1 &0&0 \cr 0&e^{i \alpha}&0 \cr 0&0&e^{i \beta}}
\label{eq:Vunitary}
\end{equation}
where $R_{ij}(\theta_{ij})$ represents a rotation by $\theta_{ij}$ in the $ij$ plane.
We have chosen a sequence of rotations which
frequently arises in the  diagonalization of simple hierarchical forms for the neutrino mass matrix, 
as illustrated in section~\ref{models}.
From equation (\ref{eq:SE}) we see that the phases $\alpha$ and $\beta$  never appear in
oscillation phenomena, and hence can be dropped, giving 
\begin{equation}
V = \pmatrix{ c_{12} c_{13} & c_{13}s_{12}&s_{13} \cr 
-c_{23}s_{12}e^{i\phi} - c_{12}s_{13}s_{23} & c_{12}c_{23}e^{i \phi} - 
s_{12}s_{13}s_{23} & c_{13}s_{23} \cr 
s_{23}s_{12}e^{i\phi} - c_{12}c_{23}s_{13} & -c_{12}s_{23}e^{i \phi} - 
c_{23}s_{12}s_{13} & c_{13}c_{23}}.
\label{eq:V}
\end{equation}
Each $R_{ij}$ must diagonalize a symmetric $2 \times 2$ sub-matrix
determining $\tan 2 \theta_{ij}$, hence,
without loss of generality, we may choose $0 \le \theta_{ij} 
\le \pi/2$, while $0 \le \phi < 2\pi$. A more convenient choice is to
keep $\theta_{12,13}$ in the first quadrant,
while $ 0 \le\theta_{23}, \phi \le \pi$.
We choose to order the neutrino mass eigenstates so that
$\Delta m^2_{23}>\Delta m^2_{12}>0$, where $\Delta m^2_{ij}\equiv m^2_i-m^2_j$.
Notice that with this parametrization $V_{e3}\ll1$ means $\theta_{13}$ close to 0 or to $90\degree$.

To study solar neutrinos,  we are interested only in the electron neutrino survival  probability,
$P_{ee}$, and hence in the evolution of $\psi_e$. This  evolution does not depend on $\theta_{23}$ or
on $\phi$ --- on  substituting (\ref{eq:Vunitary}) in (\ref{eq:SE}), 
$R_{23}$ and $\phi$ can be absorbed into redefined  states $\mu'$ and $\tau'$. Hence, we have shown
quite generally that 
$P_{ee}$ depends only on four neutrino parameters:
$\Delta m^2_{12}$, $\Delta m^2_{23}$, $\theta_{12}$ and $\theta_{13}$.

For an oscillation explanation of the atmospheric neutrino fluxes, 
$\Delta m^2_{23}$ is sufficiently large that it does not cause a resonance transition in the sun. In
the Landau-Zehner approximation, the evolution equation (\ref{eq:SE}) 
can be solved to give~\cite{Parke}
\begin{equation} P_{ee} = ( |V_{e1}|^2, |V_{e2}|^2, |V_{e3}|^2) 
\pmatrix{1-P &P&0 \cr P&1-P&0 \cr 0&0&1}
\pmatrix{|V_{e1}^{\rm m}|^2 \cr |V_{e2}^{\rm m}|^2 \cr |V_{e3}^{\rm m}|^2}
\label{eq:survival}
\end{equation}
where  $V_{ei}^{\rm m}$ are the mixing matrix elements in matter, and $P$ is  the transition
probability between the states at resonance:
\begin{equation}\label{eq:prob}
P=e^{-E_{\rm NA}/E}\theta(E-E_{\rm A}),\qquad
E_{\rm NA}=\frac{\pi\Delta m^2_{12}\sin^2(2\theta_{12})}
{4|\frac{1}{N_e}\frac{d N_e}{dx}|_1\cos(2\theta_{12})},\qquad
E_{\rm A}=\frac{\Delta m^2_{12} \cos 2\theta_{12}}{2\sqrt{2}G_{\rm F} |N_e|_0\cos^2\theta_{13}}
\end{equation}
Here $E$ is the neutrino energy, $\theta$ is the step function,
the 1 subscript indicates that $N_e$ and its
gradient $dN_e/dx$ are evaluated at the resonance point,
while the 0 subscript indicates the production point.
The large mass splitting $\Delta m^2_{23}$ enters
$P_{ee}$ only via the  matter mixing angles, and decouples from these expressions in the limit  that it
is much larger than $A_{11} E$, and also in the limit that 
$\theta_{13}$ vanishes. For most of this section we make $\Delta  m^2_{23}$ sufficiently large that it
decouples, and we comment at the end  on the effect on the allowed regions of parameter space for
non-zero 
$\theta_{13}$ and small $\Delta m^2_{23}$, where $\Delta m^2_{23}$ effects may not decouple.

\begin{figure}[t]\begin{center}
\begin{picture}(18,4.5)
\putps(0,-0.5)(0,-8.6){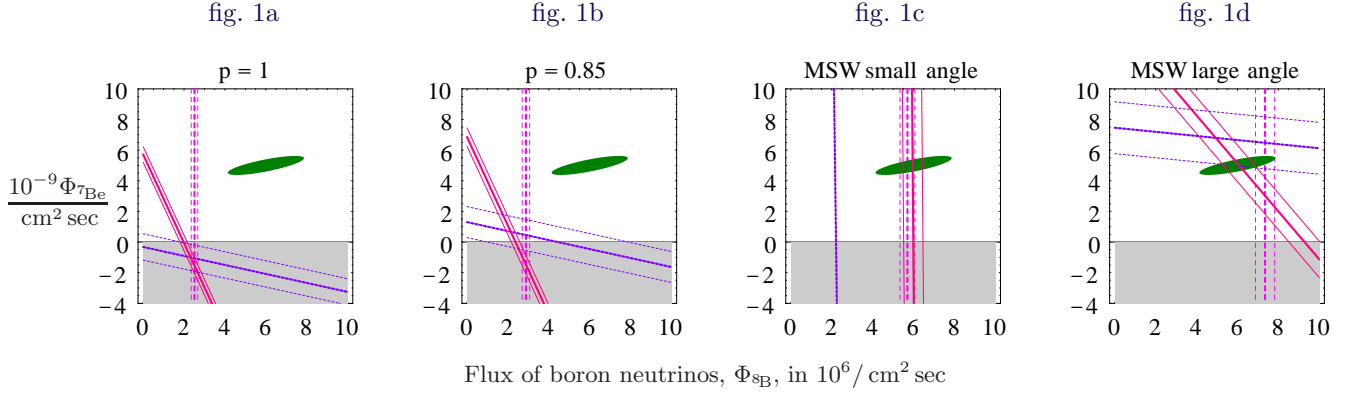}
\put(-0.6,1.5){\small$\displaystyle\frac{10^{-9}\PhiBe}{\cm^2\sec}$}
\put(5.5,-0.8){\small Flux of boron neutrinos, $\PhiB$, in $10^6/\cm^2\sec$}
\Blue
\put(2.1,4){fig.~\ref{fig:Incroci}a}
\put(6.4,4){fig.~\ref{fig:Incroci}b}
\put(10.7,4){fig.~\ref{fig:Incroci}c}
\put(15.0,4){fig.~\ref{fig:Incroci}d}\Black
\end{picture}
\vspace{0.5cm}
\caption[SP]{\em Values of $(\PhiB,\PhiBe)$ measured by
the Chlorine experiment (continuous lines), the Gallium experiment (dashed lines)
and by the SuperKamiokande experiment (long dashed lines) assuming various neutrino oscillation schemes:
$\bullet$ no oscillation in fig.~\ref{fig:Incroci}a;
$\bullet$ an energy-independent $P(\nu_e\to\nu_e)=0.85$ in fig.~\ref{fig:Incroci}b;
$\bullet$ the best-fit point of the small-angle MSW oscillation in fig.~\ref{fig:Incroci}c;
$\bullet$ the best-fit point of the large-angle MSW oscillation in fig.~\ref{fig:Incroci}d.
\label{fig:Incroci}}
\end{center}\end{figure}

The signals $S_i$ at the three types of solar neutrino experiments are
\begin{equation} S_i =
\int\!dE\,\Phi(E)\big[\sigma_i^e(E)P_{ee}(E)+\sigma_i^{\not\, e}(E)\bigl(1-P_{ee}(E)\bigr)\big],
\quad i = \{{\rm SK, Ga, Cl}\}
\label{eq:sig}
\end{equation} where $\Phi(E)$ is the total flux of solar neutrinos with energy $E$, and
$\sigma_i^{e,\not\, e}(E)$ are the interaction cross sections at experiment $i$ for electron-type and
non-electron-type  neutrinos, respectively (only the water Cerenkov detectors are sensitive to neutral
currents, so $\sigma_{\rm Ga}^{\not\, e}(E) = \sigma_{\rm Cl}^{\not\, e}(E) = 0$).   We will use the theoretical predictions
of the various cross sections found in~\cite{BahcallBook,BahcallWWW}.  The flux $\Phi(E)$ is broken into
components in the standard way by specifying the  production reaction, giving~\cite{BahcallBook}
\begin{equation} \Phi(E) = \sum_\alpha \Phi_\alpha f_\alpha(E),\qquad\hbox{with}\qquad
\int_0^\infty f_\alpha(E)\,dE=1
\label{eq:flux}
\end{equation}
and $\alpha = \rm pp, p\hbox{$e$}p, ^7\!Be,^{13}\!N,^{15}\!O,{}^{17}\!F,{}^8\!B,h\hbox{$e$}p$.
At this point we follow the (nearly) model-independent treatment of the fluxes described
in~\cite{CDFLR} by making the following assumptions:
\begin{enumerate}
\item The energy dependence $f_\alpha(E)$ of the single components of the neutrino fluxes
predicted by solar models (\cite{BahcallBook,BP} for instance) are correct.
In fact the $f_\alpha(E)$ do not depend on the structure of the sun,
and are the same in any solar model that does not introduce non-standard electroweak effects~\cite{BahcallBook}.

\item The overall $\Phi_\alpha$ can differ from their solar models predictions.
However there are strong physical reasons to believe that the ratios
$\Phi_{^{13}\!\rm N}/\!\Phi_{^{15}\!\rm O}$ and $\Phi_{{\rm p}e{\rm p}}/\!\Phi_{\rm pp}$ can be set to
their solar SM values~\cite{BP}.
Furthermore we neglect entirely
hep and $^{17}$F neutrinos, which we expect to be extremely rare.

\item The present total luminosity of the sun, $K_\odot$, determines its present total neutrino luminosity as
\begin{equation} K_\odot =
\sum_\alpha \left(\frac{Q}{2} - \langle E_{\nu_\alpha} \rangle \right)\Phi_\alpha\approx
\frac{Q}{2}\sum_\alpha\Phi_\alpha, 
\label{eq:sl}
\end{equation}
where $Q = 26.73$ MeV is the energy released in the reaction $4{\rm p}  + 2e \rightarrow 
\,^4\!{\rm He} + 2\nu_e$, and $K_\odot = 8.53\cdot 10^{11}$ MeV cm$^{-2}$ s$^{-1}$
is the solar radiative flux at the earth.
Using (\ref{eq:sl}) amounts to assuming that the
solar energy comes from nuclear reactions that reach completion,  and that the sun
is essentially static over the $10^4$ years employed by photons to random-walk out of the solar interior.   
\end{enumerate}
After the first assumption we have one free parameter $\Phi_\alpha$ for each
$\alpha$; the second then reduces the number of free parameters to four, which we  can take to be
\begin{equation}
\Phip\equiv \Phi_{\rm pp}+\Phi_{{\rm p}e{\rm p}},\qquad
\PhiCNO\equiv \Phi_{^{13}\!\rm N}+\Phi_{^{15}\!\rm O},\qquad
\PhiBe\qquad\hbox{and}\qquad\PhiB.
\end{equation}
The luminosity constraint  allows us to  eliminate $\Phip$, giving
\begin{equation} S_i = S_i(\Delta m_{12}^2, \,\theta_{12}, \,\theta_{13};
\PhiB,\PhiBe,\frac{\PhiCNO}{\PhiBe}).
\label{eq:sigdep}
\end{equation}
Since solar models give a stable prediction for
$\PhiCNO/\PhiB=0.22$~\cite{CDFLR},
we have singled out this ratio and we will use its SSM value in our analysis.
Variations of even an order of magnitude in the ratio affect negligibly our final results,
since the two neutrino components have similar cross sections in existing detectors.

\begin{figure}[t]\begin{center}
\begin{picture}(8,6)
\putps(0,0)(0,0){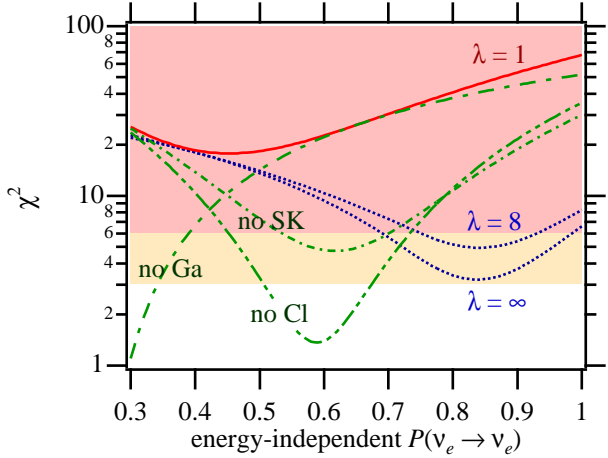}
\end{picture}\hspace{1.3cm}\raisebox{2cm}{
\parbox[b]{6cm}{\caption[c]{\em Values of the $\chi$-square
as function of an energy independent $P(\nu_e\to \nu_e)$.
The parameter $\lambda$ is defined in eq.\eq{chiq}.
Also shown is the $\chi^2$ with one experiment discarded and $\lambda=1$.
\label{fig:chiq}}}}
\end{center}\end{figure}

\begin{figure}[t]\begin{center}
\begin{picture}(18,11)
\putps(0,-0.5)(-3.6,-0.5){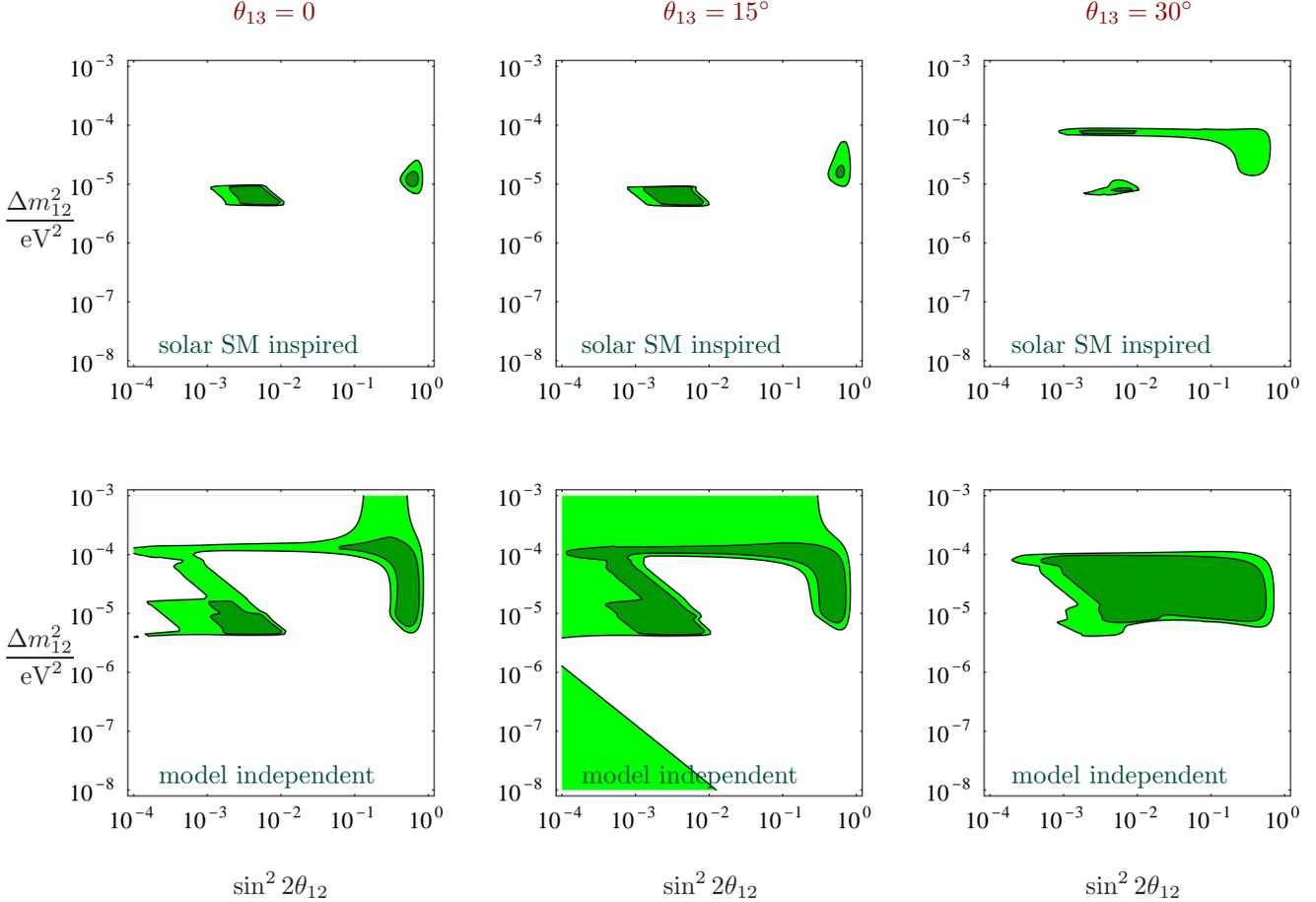}
\put(-0.4,2.3){$\displaystyle{\Delta m^2_{12}\over\eV^2}$}
\put(-0.4,8.2){$\displaystyle{\Delta m^2_{12}\over\eV^2}$}
\put(2.7,-0.8){$\sin^22\theta_{12}$}
\put(8.5,-0.8){$\sin^22\theta_{12}$}
\put(14.2,-0.8){$\sin^22\theta_{12}$}\Red
\put(2.7,11){$\theta_{13}=0$}
\put(8.5,11){$\theta_{13}=15\degree$}
\put(14.2,11){$\theta_{13}=30\degree$}\Green
\put(1.7,6.5){solar SM inspired}
\put(7.4,6.5){solar SM inspired}
\put(13.2,6.5){solar SM inspired}
\put(1.7,0.7){model independent}
\put(7.4,0.7){model independent}
\put(13.2,0.7){model independent}\Black
\end{picture}
\vspace{0.5cm}
\caption[SP]{\em Allowed regions in the plane $(\sin^2 2\theta_{12},\Delta m^2_{12})$
for $\theta_{13}=0,15\degree$ and $30\degree$.
The upper plots assume that the BP solar model is correct.
The lower plots are the result of the solar model independent analysis described in the text.
\label{fig:Sun}}
\end{center}\end{figure}

\begin{figure}[t]\begin{center}
\begin{picture}(6,5.5)
\putps(0.8,0.5)(0.8,-10.5){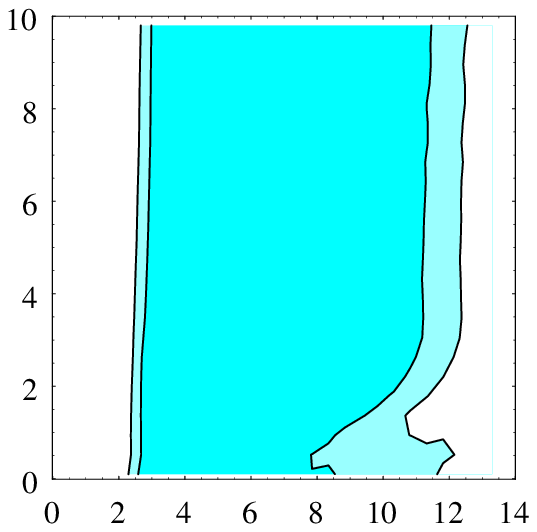}
\putps(0,0)(0,0){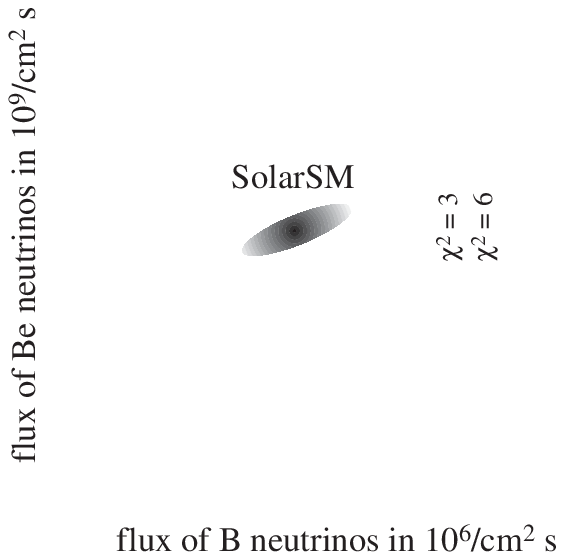}
\end{picture}\hspace{1.3cm}\raisebox{3cm}{
\parbox[b]{6cm}{\caption[c]{\em Isoplot of $\chi^2_8$, minimized in the mixing parameters.
\label{fig:Phi}}}}
\end{center}\end{figure}

\subsection{Model-independent solar analysis --- all experiments}\label{SSMindep}
The signals now depend only on $\PhiB$ and
$\PhiBe$, so that, for any given oscillation pattern $P_{ee}(E)$
it is possible to plot the three experimental results\footnote{The SuperKamiokande
experimentalists give directly the value of the flux they measure.
The other experiments involve more uncertain neutrino cross sections and prefer to
give the frequency of events measured per target atom in their detector.
For simplicity we have omitted this detail in the text,
leaving a trivial inconsistency
between eq.\eq{sksig} and\eq{sig}.}~\cite{SK-solar,ClSun,GaSun,KaSun}
\begin{eqnsystem}{sys:S}
S_{\rm Cl}^{\rm exp} &=& (2.54\pm 0.20)\,\,{10^{-36} {\rm s}^{-1}}\label{eq:clsig}\\
S_{\rm Ga}^{\rm exp} &=& (75 \pm 7)\,\,{10^{-36} {\rm s}^{-1}} \label{eq:gasig}\\
S_{\rm SK}^{\rm exp} &=& (2.51\pm0.16)\cdot 10^6 \,{\rm cm}^{-2}{\rm s}^{-1}\label{eq:sksig}
\end{eqnsystem}
as three bands in the ($\PhiB$, $\PhiBe$) plane.
The three bands will in general not meet,
giving interesting solar model independent restrictions
on the oscillations parameters.

\smallskip

We begin the analysis by studying the case of {\em no neutrino oscillations\/} ($P_{ee}=1$).
In this particular case the solar model independent analysis does not give a strong result.
Surprisingly the three bands perfectly meet~\cite{CDFLR,Langacker} as shown in fig.~\ref{fig:Incroci}a,
but mainly in the unphysical
$\PhiBe<0$ region, with a small area in the physical region
lying within $2\sigma$ of each central value.
Since the physical crossing region has a negligible $^{7}\!$Be flux,
the value of $\PhiCNO/\PhiBe$ becomes completely irrelevant.

To discuss this case in a quantitative way and to deal with more general cases it is useful to introduce
the $\chi$-square function
\begin{equation}
\chi^2_{\lambda}\left(P_{ee}(\Delta m_{12}^2, \,\theta_{12}, \,\theta_{13}), \,\PhiB, \,\PhiBe\right)
\equiv \sum_i\left(\frac{S_i-S_i^{\rm exp}}{\Delta
S_i^{\rm exp}}\right)^2+\sum_{jk}\frac{(\Phi_j-\Phi_j^{\rm SSM})(\Phi_k-\Phi_k^{\rm SSM})}{\lambda^2 \Delta \Phi_{jk}^{2\,\rm SSM}}
\label{eq:chiq}
\end{equation}
where $\Delta S_i^{\rm exp}$ is the $1\sigma$ uncertainty for experiment $i$, given
in~(\ref{sys:S}),
$\Phi^{\rm SSM}$ is the flux prediction of the solar model~\cite{BP} and $\Delta
\Phi^{\rm SSM}$ is the corresponding error matrix, taken with some generosity.
The $1\sigma$ ranges of $\PhiB$ and $\PhiBe$ are represented by the ellipse in fig.~\ref{fig:Incroci}.
We perform our analysis with two choices for
$\Delta \Phi=\lambda\cdot \Delta \Phi^{\rm SSM}$.
We call the first choice, $\Delta \Phi=\Delta \Phi^{\rm SSM}$,
``{\em solar SM inspired\/}''). 
The second choice, $\Delta \Phi=8\cdot\Delta \Phi^{\rm SSM}$  (``{\em model independent\/}'')
has the same shape as the first, but is eight times as large.
The part of the analysis done using this $\Delta \Phi$ is virtually free of solar physics input.
The choice $\lambda=8$ (rather than $\lambda=\infty$) avoids unnatural values of $\PhiBe$.
This choice essentially ignores solar physics considerations, but the virtue of
having a number of independent experimental results is precisely that we need no longer rely heavily on
solar modelling to gain insight into the underlying particle physics.

Minimizing the $\chi^2$ in the positive flux region we obtain $\min\chi^2_8(P_{ee}=1)=8.25$.
The usual criterion for goodness of fit says that a $\chi^2$ with one
degree of freedom larger than $8.25$  is obtained with a very small probability, $\wp\approx 0.4\%$
(a careful Monte Carlo treatment of the $\PhiBe>0$ constraint gives similar results~\cite{Langacker}).
We however remark that, if the sun really emits the best-fit fluxes,
$\PhiBe=0$ and $\PhiB=2.5~10^6/{\rm cm}^2{\rm s}$,
there is a 10\% probability that statistical fluctuations produce the present experimental data.

\medskip

We can just as easily investigate the slightly more general case of an energy independent $P_{ee}$.
The dependence on the
neutrino parameters $\Delta m_{12}^2$, $\theta_{12}$, and $\theta_{13}$ arises through $P_{ee}$; if the
survival probability is a constant,  then we can minimize $\chi^2_\lambda$ in the positive-flux region for any
value of $P_{ee}$ to obtain $\min\chi_{\lambda}^2(P_{ee})$, which is plotted in fig.~\ref{fig:chiq}
for $\lambda=1$ (SSM analysis), $\lambda=8$ (SSM independent analysis) and
$\lambda=\infty$ (completely model independent analysis).
For $P_{ee} \sim0.85$, $\min\chi_{8}^2$ drops to 5,
but the fluxes required to get relatively small $\chi^2$ values
are disfavoured by solar physics considerations ---
$\PhiCNO$ and $\PhiBe$ must be nearly made to
vanish, as shown in fig~\ref{fig:Incroci}b. 
When $P_{ee}\circa{<}1/2$ the (accidental?) threefold crossing no longer occurs,
so that this case can be firmly excluded in a solar-model independent way~\cite{CDFLR,PeeCteBad}
(see fig.~\ref{fig:chiq}).
However, as we shall see in subsection~\ref{OneExpWrong},
once we allow for the possibility that one type of experiment's
results should be discarded, it is possible to obtain good fits of the data for constant $P_{ee}\sim1/2$
without having to resort to unnatural flux values.

\medskip

Of course, we are interested in any points in parameter space that fit the data well, regardless of
whether they lead to constant $P_{ee}$.  For any values of $\Delta m_{12}^2$, $\theta_{12}$, and
$\theta_{13}$ we can make plots similar to fig.~\ref{fig:Incroci}a.
Fig.s~\ref{fig:Incroci}c and \ref{fig:Incroci}d show two examples that illustrate
the familiar 2-generation small and large angle MSW solutions, which evidently fit the data well if
standard solar model fluxes are used.

In fig.~\ref{fig:Sun} we show how the allowed regions in neutrino
parameter space change if we let the fluxes vary over an expanded range of values.
For each point in ($\Delta m_{12}^2$, $\theta_{12}$, $\theta_{13}$) space, we minimize $\chi^2_1$
and $\chi^2_8$ by varying the fluxes within the physical region, and then we plot
contours of $\min\chi^2_{\lambda}$ in the 
($\sin^2(2\theta_{12})$, $\Delta m_{12}^2$) plane for various values of
$\theta_{13}$. 
The results for the ``SSM inspired'' and ``model independent'' analyses are shown in fig.s~\ref{fig:Sun}
(upper row and lower row, respectively).
The contours are for $\chi^2=3$ and $\chi^2=6$.

For small $\theta_{13}$ the ``{\em SSM inspired\/}'' results show the standard small and large angle MSW regions. 
For larger values of $\theta_{13}$, the two MSW regions join, and, as $\theta_{13}$ approaches $\pi/4$,
the solutions with large $\theta_{12}$ disappear.
For $\theta_{13}=\pi/4$ the region with $\min\chi_{1}^2<3$ is in fact absent entirely.

The ``{\em model independent\/}'' results similarly exhibit a very strong $\theta_{13}$ dependence.
We see that the ``model independent'' analysis continues to give strong restrictions of the
oscillation parameters --- in particular the $\Delta m_{12}^2$ values with
$\min\chi^2_{8}<3$ are always in the range $\sim 10^{-(4\div 5)}\eV^2$.
This will not remain true when we
consider the consequences of ignoring one experiment's data in subsection~\ref{OneExpWrong}.

If $\PhiCNO/\PhiB$ is ten times larger than in SSM there are new allowed regions.
However these possible new regions, with $\Delta m^2=10^{-(5\div 6)}\eV^2$ and
$\sin^2 2\theta_{12}\circa{>}10^{-2}$, are excluded in a model-independent way
by the non observation of a day/night asymmetry at SuperKamiokande~\cite{SK-solar,RecentSunAnalysis}.
The recent data~\cite{SK-solar} on this asymmetry in fact disfavour
as well the large angle MSW
solution of the SSM-inspired analysis.
Moreover, we have not included in our $\chi^2$ analysis the
SuperKamiokande measurement of the distortion of the $^8\!$B spectrum~\cite{SK-solar,RecentSunAnalysis},
because the present positive $1\sigma$ signal could be
produced by a $\Phi_{{\rm h}e\rm p}/\PhiB$ ratio 15 times
larger than the prediction of BP95~\cite{BP}.
Without a very large h$e$p flux, the present measurement excludes an otherwise allowed region with
$\Delta m^2\approx 10^{-4}\eV^2$
and $\sin^22\theta_{12}$ in the range $10^{-4}\div 10^{-1}$~\cite{Langacker,Petcov}.

\medskip

Our model independent analysis allows us to investigate how well present experiments
are able to measure the SSM-independent neutrino fluxes $\PhiB$ and $\PhiBe$.
This question is answered in fig.~\ref{fig:Phi}, where we plot the values of the fluxes
that can give a good ($\chi^2_8<6$) or very good ($\chi^2_8<3$) fit for
{\em some} value of the oscillation parameters
$\Delta m^2_{12}$, $\theta_{12}$ and $\theta_{13}$.
We see that the value of $\PhiB$ is currently determined with
an error larger than the solar model expectation.
It will be directly measured in the new on-going SNO experiment.
On the contrary the value of $\PhiBe$ is at present totally unknown:
in fact in the small angle MSW solution the monochromatic $^{7}\!\rm Be$ flux can be completely converted
into $\not\! e$ neutrinos,
that are not detected by existing experiments.
Borexino will be able to detect neutral currents effects in this range of energies
and probably allow a direct determination of $\PhiBe$~\cite{Langacker}.

\medskip

As discussed above we perform our analysis under the assumption that $\Delta m_{23}^2$
is large enough that its effects decouple. 
For any given $\Delta m_{23}^2$ it is straightforward to
reproduce fig.~\ref{fig:Sun}
by using the exact expressions for $\theta_{12}^{\rm m}$
and $\theta_{13}^{\rm m}$ in equation (\ref{eq:survival}).
In this way we find that for small $\theta_{13}$
($\circa{<}15\degree$),  our results are insensitive to $\Delta m_{23}^2$ down to $\Delta
m_{23}^2=5 \cdot 10^{-4}\eV^2$.  For large $\theta_{13}$, $\Delta m_{23}^2$ effects start to become
noticeable when $\Delta m_{23}^2$ drops below $2 \times10^{-3}\eV^2$; for example, for $\theta_{13}=
40\degree$ and $\Delta m_{23}^2=5 \cdot 10^{-4}\eV^2$, the allowed region in the SM inspired
analysis is significantly smaller than in the decoupled limit, with the $\chi^2_{\rm min}<6$ region never
reaching $\sin^2(2\theta_{12})> 0.1$.  In spite of these changes for small $\Delta m_{23}^2$, the
essential features of fig.~\ref{fig:Sun} in any case remain unchanged.

\begin{figure}[p]\begin{center}
\begin{picture}(17,15)
\putps(0,-0.5)(-3.6,-0.5){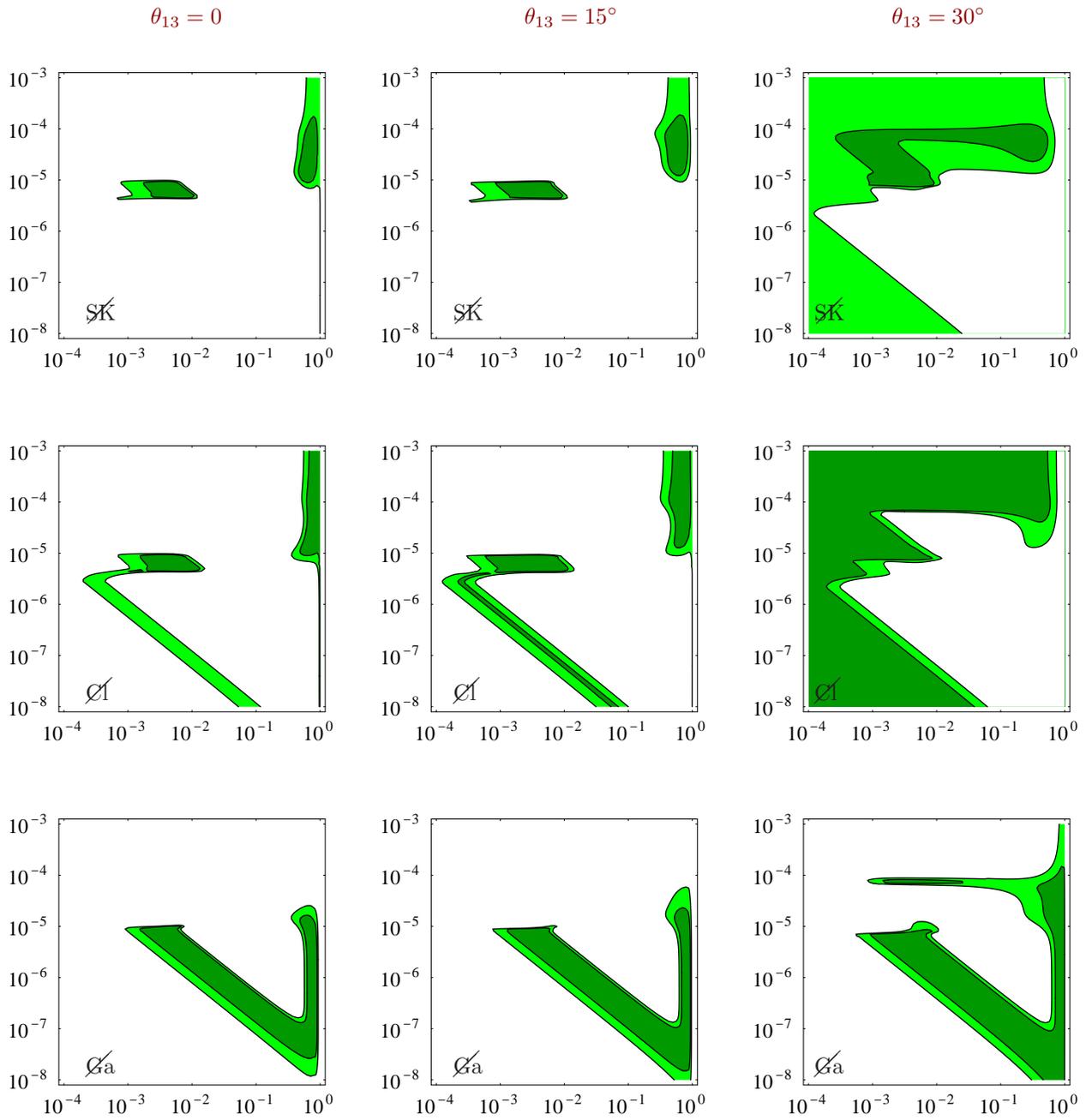}\Red
\put(2.7,17){$\theta_{13}=0$}
\put(8.5,17){$\theta_{13}=15\degree$}
\put(14.2,17){$\theta_{13}=30\degree$}\Black
\put(1.7,12.4){\noSK}\put(7.4,12.4){\noSK}\put(13.0,12.4){\noSK}
\put(1.7,6.5){\noCl} \put(7.4,6.5){\noCl} \put(13.0,6.5){\noCl}
\put(1.7,0.7){\noGa} \put(7.4,0.7){\noGa} \put(13.0,0.7){\noGa}
\end{picture}
\vspace{0.5cm}
\caption[SP]{\em fits of the solar data in the plane $(\sin^2 2\theta_{12},\Delta m^2_{12}/\eV^2)$
for $\theta_{13}=0$, $15\degree$ and $30\degree$
assuming that one of the three solar neutrino experiments has a large unknown systematic error
(SuperKamiokande in the first row, Chlorine in the second and Gallium in the third)
and is therefore discarded from the analysis.
The contours are for $\chi^2=3$ and $\chi^2=6$.
\label{fig:SunExpLess}}
\end{center}\end{figure}

\subsection{Model independent solar analysis --- one experiment ignored}\label{OneExpWrong}
In subsection~\ref{SSMindep}, the present level of experimental evidence allowed us to omit
one restriction (the solar model) and still yield interesting results.
Likewise, we can choose to omit one experiment from the analysis while
keeping some solar information and still yield interesting results.

The motivation for this is obvious: neutrino experiments are extremely
difficult to perform and particular detection schemes may suffer from
some systematic error previously not considered. We make no judgements
here about the errors associated with any particular experiment. Instead
we consider analyses where we do not include one class of experiment, either
water-Cerenkov, gallium or chlorine, which we designate $\noSK$, $\noCl$
and $\noGa$ respectively. However, because we are losing an experiment, it is impossible
to analyze the data without some level of information regarding the solar
model. Consequently, we perform the analysis only within the solar SM inspired region.
The results for this analysis are shown in
figures~\ref{fig:SunExpLess}
(upper row: without SuperKamiokande data,
middle row: without chlorine data, and
lower  row: without gallium data).

The $\noSK\,$ case largely resembles the complete data set analysis, with
some additional space allowed in the higher $\Delta m^2_{12}$ region. In
contrast, the other two cases ($\noGa\,$ and $\noCl$) show considerable
differences.

For the $\noGa\,$ case, there is a strong preference for either small
$\Delta m^2_{12}$ or large $\theta_{12}$ and $\theta_{13}$. For the
$\noCl$ case, for both large $\theta_{12}$ and $\theta_{13}$ we have
the presence of large regions with large $\Delta m^2_{12} = 10^{-4}\eV^2$,
above the level-crossing threshold, and with small $\Delta m^2_{12}$, in the
non-adiabatic region.  In either case, in a large portion of these regions
matter enhancements are unimportant. That is, in the absence of one of these
two classes of experiment, given sufficiently large angles, the solar neutrino
 problem can be resolved {\it simply by vacuum oscillations alone!} In such a
case, new experiments, such as Borexino, would see an absence of energy
dependence in the electron neutrino survival probability.

\section{Atmospheric and Solar Neutrinos: The Minimal Scheme}\reseteq\label{AtmStandard}
The simplest picture for reconciling both solar and atmospheric
neutrino fluxes via oscillations of $\nu_{e,\mu,\tau}$ results when
there is a  hierarchy $|\Delta m^2_{23}| \gg |\Delta m^2_{12}|$,
and $\Delta m_{12}^2$ is too small to affect oscillations of atmospheric
neutrinos.
In section~\ref{Sun}, we showed that in this case the solar fluxes depend only on
$\Delta m^2_{12}$, $\theta_{12}$ and
$\theta_{13}$\footnote{Although for non-zero $\theta_{13}$, there is a dependence on $\Delta
m^2_{23}$ if it is small enough.}, and below we show that the atmospheric fluxes depend only on
$\Delta m^2_{23},
\theta_{23}$ and $\theta_{13}$. In the limit that $\theta_{13}=0$, the two phenomena become
independent, in the sense that they depend on no common parameters: solar oscillations are
$\nu_e \rightarrow \nu_\mu$ at a low frequency, while atmospheric oscillations are $\nu_\mu
\rightarrow
\nu_\tau$ at a much higher frequency. However, solar oscillations are allowed for a wide range of
parameters with large $\theta_{13}$, and the atmospheric data does not require $\theta_{13}$ to be
very small. Hence, in this section we explore this simple picture keeping $\theta_{13}$ as a free
parameter. We comment on the alternative possibility --- that $\Delta m^2_{12}$ is large enough to
contribute to atmospheric neutrino oscillations --- in section~\ref{AtmNonStandard}.

Matter
effects in the earth are important only for a relatively small fraction
of the atmospheric neutrinos, those with high energy,
and they are neglected here\footnote{For more details see e.g.\ ref.~\cite{LL}.}.
In this case, (\ref{eq:SE}) can be integrated to give oscillation
probabilities $P_{ff'}(t) = |A_{ff'}(t)|^2$, where $A$ is given by the
matrix equation
\begin{equation}
A(t) = V e^{-iEt} V^\dagger.
\label{eq:A}
\end{equation}
Since an overall phase in $A$ is irrelevant to $P$, and  $\Delta
m^2_{12}$ effects are negligible, we may make the substitution
\begin{equation}
e^{-iEt} \longrightarrow {\rm diag}\,(1,1, e^{-i \Delta m_{23}^2 t/2E})
\label{eq:sub}
\end{equation}
Using the form (\ref{eq:Vunitary}) for $V$, we immediately discover
that the probabilities are independent of $\theta_{12}$ and $\phi$, as
well as $\alpha$ and $\beta$. The probabilities are given by
\begin{eqnsystem}{sys:PatmStandard}
P_{e \mu}    & = & s_{23}^2 \sin^2 2 \theta_{13}   \; S_{23}  \\
P_{e \tau}   & = & c_{23}^2 \sin^2 2 \theta_{13}   \; S_{23} \\
P_{\mu \tau} & = & c_{13}^4 \sin^2 2 \theta_{23}   \; S_{23} \label{eq:P}\\
\riga{or equivalently, by unitarity}\\[-3mm]
P_{ee}       & = & 1 - \sin^2 2 \theta_{13} \; S_{23}  \\
P_{\mu \mu}  & = & 1 - 4 c_{13}^2 s_{23}^2 (1-c_{13}^2 s_{23}^2)\;S_{23}  \\
P_{\tau \tau}& = & 1 - 4 c_{13}^2 c_{23}^2 (1-c_{13}^2 c_{23}^2)\; S_{23}
\label{eq:Pdiag}
\end{eqnsystem}
where $S_{23} = \sin^2 (\Delta m^2_{23}t/4E)$.
The parameter $\Delta m^2_{23}$ can be extracted from the data by
fitting to the zenith angle distribution of the events. Here we
concentrate on the determination of the parameters $\theta_{13}$ and
$\theta_{23}$. These can be extracted, independent of the value of
$\Delta m^2_{23}$, if we assume that the downward going neutrinos have not
oscillated, while the upward going neutrinos are completely
oscillated, so that $S_{23}$ is averaged to 0.5.
In view of the reported angular distribution of the multi-GeV data for
1-ring $e$-like, 1-ring $\mu$-like and partially contained (PC) events
\cite{SK-atm}, this assumption appears to be valid, at least for
angular cone sizes about the vertical which are not too large.
For events of class $i$, which are induced by $\nu_e$ charged current,
$\nu_\mu$ charged current and neutral current interactions with
relative probabilities $f_{eCC}^i, f_{\mu CC}^i$ and $f_{NC}^i$,
the up-down ratio $\rho_i$ is given by
\begin{equation}
\rho_i = {N_i^\uparrow \over N_i^\downarrow} =
f_{eCC}^i \cdot (P_{ee} + r P_{e \mu }) + f_{\mu CC}^i \cdot( P_{\mu \mu}
+ { 1 \over r} P_{e \mu}) + f^i_{NC}.
\label{eq:ratio}
\end{equation}
where we have set $S_{23} = 0.5$, and
$N_i^{\uparrow, \downarrow}$ are the number of upward and
downward events of class $i$. We are interested in $i$ being 1-ring
$e$-like, 1-ring $\mu$-like and PC. The overall normalization of these
event numbers has considerable uncertainties due to the calculation of
the neutrino fluxes produced in cosmic ray showers, hence we consider
three up-down ratios
\begin{eqnsystem}{sys:atmexp}
\rho_e    & = & 1.23 \pm 0.29   \\
\rho_\mu  & = & 0.62 \pm 0.16   \\
\rho_{PC} & = & 0.48 \pm 0.12\label{eq:ratioexp}\\
\riga{and two ratios of downward going fluxes}\\[-2mm]
{N_\mu^\downarrow + N_{PC}^\downarrow \over N_e^\downarrow}
=  \xi r  &=& 3.0 \pm 0.6 \label{eq:Nd}  \\
{N_{PC}^\downarrow \over N_\mu^\downarrow} =
\xi' & = & 1.3 \pm 0.3 \label{eq:Ne}
\end{eqnsystem}
where $r$ is the ratio of $\nu_\mu$ to $\nu_e$ fluxes.
The numbers give the
Super-Kamiokande data, extracted from the figures of Ref.~\cite{SK-atm},
with upward and downward directions defined by the azimuthal angle
having $\cos \theta$ within 0.4 of the vertical direction.
The parameters $\xi r$ and $\xi'$ represent the theoretical values for the
ratios of (\ref{eq:Nd}) and (\ref{eq:Ne}). These two downward going ratios
do not involve oscillations, and the Super-Kamiokande collaboration compute
Monte Carlo values of 3.1 and 1.0, respectively, agreeing very well with the
data. Since these two ratios do not probe oscillations, at least within our
assumptions, we do not use them for the fits below. We do not
use the sub-GeV data as the poor angular correlation between the
neutrino and charged lepton directions leads to a smoothing of the
up-down ratio.
From the flux calculations of Honda {\it et al}~\cite{Honda}, and
using the measured momentum distributions for the events~\cite{SK-atm}, we
estimate $r= 4.0 \pm 0.5$, for this multi-GeV data near the vertical
direction. A more refined analysis would use a larger value of $r$ for
PC events than for FC events.

\begin{figure}[t]\begin{center}
\begin{picture}(18,10)
\putps(0,6.5)(0,5.35){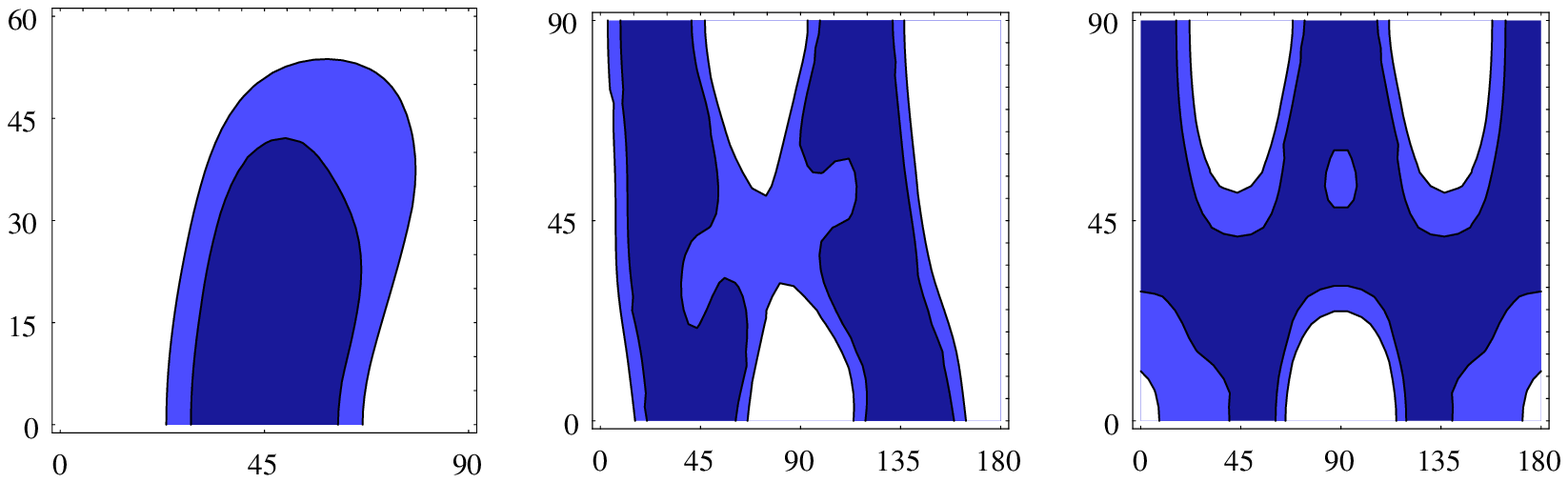}
\putps(0,0.5)(0,-0.65){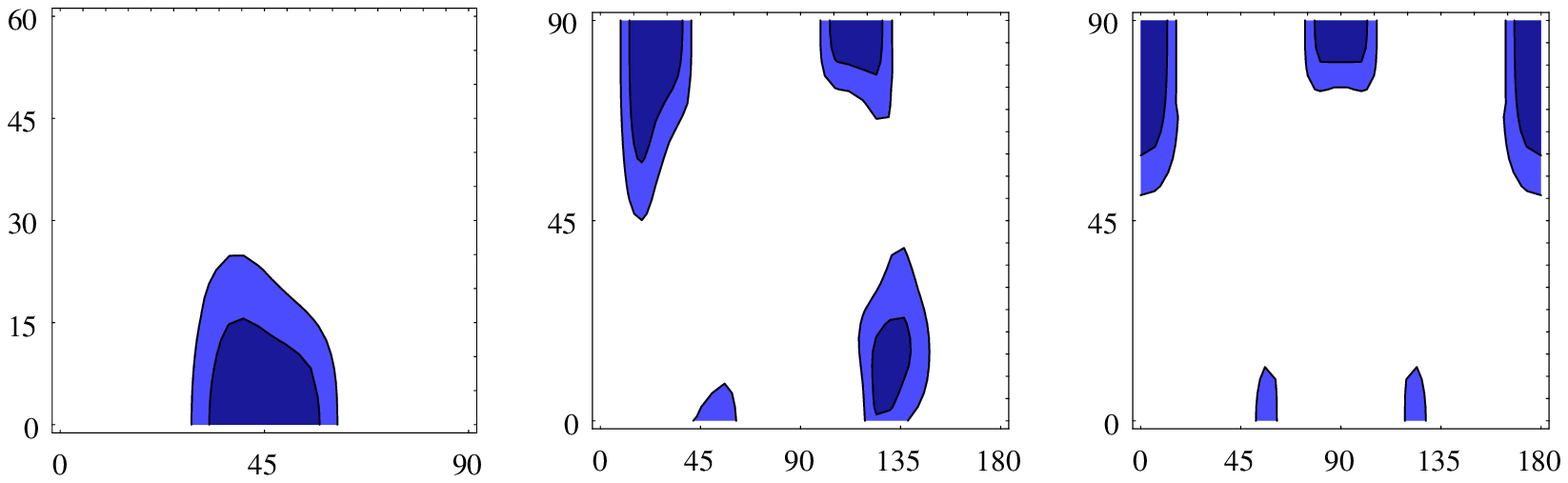}
\put(0,2.4){$\theta_{13}$}
\put(0,8.4){$\theta_{13}$}
\put(3.2,-0.8){$\theta_{23}$}\put(8.8,-0.8){$\theta_{23}$}\put(14.4,-0.8){$\theta_{23}$}
\puttag(3.2,11){(a)}  \puttag(8.9,11){(b1)}  \puttag(14.7,11){(b2)}
\puttag(3.2,5){(a$'$)}  \puttag(8.9,5){(b1$'$)}  \puttag(14.7,5){(b2$'$)}

\end{picture}
\vspace{0.5cm}
\caption[SP]{\em Mixing angles $\theta_{ij}$ that fit
the up/down ratios~{\rm (\ref{sys:atmexp}a,b,c)}
of atmospheric neutrinos, assuming that
\tag{(a)} $\Delta m^2_{12}\ll 10^{-3}\eV^2$ and any $\theta_{12}$,
\tag{(b)} $\Delta m^2_{12}\approx \Delta m^2_{23}\approx 10^{-3}\eV^2$, $\phi=0$ and
\tag{(b1)} $\theta_{12}=20\degree$,
\tag{(b2)} $\theta_{12}=45\degree$.
Primed figures are as above, but including in the asymmetry also the intermediate bins
in the angular distribution of~\cite{SK-atm} (see text).
The contours are for $\chi^2=3$ and $\chi^2=6$.
\label{fig:Atm}}
\end{center}\end{figure}

The results of a fit of the three up/down ratios to the two free parameters
$\theta_{23}$ and $\theta_{13}$ are shown in figure~\ref{fig:Atm}(a).
We have obtained the fractions $f^i_{eCC, \mu CC, NC}$ from the
Monte Carlo results of the Super-Kamiokande collaboration
\cite{SK-atm}, and we have used the oscillation probabilities of~(\ref{eq:P}).
In order to work with Gaussian distributed experimental data,
we have directly fitted the six measured neutrino numbers
$N_i^{\uparrow,\downarrow}$
leaving arbitrary the three overall fluxes of each type,
$N_i^\uparrow+N_i^\downarrow$.
The preferred region of the plot is easy to understand,
since at the point $\theta_{23} = 45\degree$ and $\theta_{13}=0$, the
$\nu_e$ are unmixed, while there is complete $\nu_\mu \leftrightarrow
\nu_\tau$ mixing, so $\rho_e \simeq 1$ and $\rho_\mu \simeq \rho_{PC} \simeq
0.5$.
It is apparent from Fig. \ref{fig:Atm}(a) that this minimal scheme is
allowed for a large range of angles about this point: $\theta_{23} =
45\degree \pm 15\degree$ and $\theta_{13} = 0 \div 45\degree$.

If the solar neutrino fluxes, measured by all three techniques,
are to agree with solar model inspired values, then the results of
section~\ref{Sun} show that $\Delta m_{12}^2$
is too small to affect atmospheric oscillations, it is either of order
$10^{-4} - 10^{-5}\eV^2$ or of order $10^{-10}\eV^2$. In this
case, the minimal scheme for atmospheric neutrinos, described in this
section, is the unique possibility using just the three known
neutrinos. This observation enhances the importance of the fit of
figure~\ref{fig:Atm}(a); further data will reduce the allowed region,
as the three up-down ratios of (\ref{eq:ratioexp}) have
small systematic uncertainties and are statistics limited.
The solar neutrino fluxes do not put extra constraints on
the value of $\theta_{13}$, although it becomes correlated with
$\theta_{12}$, as shown in figure~\ref{fig:Sun}.
If the atmospheric flux measurements
require $\Delta m^2_{23} > 2 \times 10^{-3}\eV^2$, then the limit on
$P_{ee}$ from the CHOOZ experiment~\cite{CHOOZ} requires $\theta_{13} < 13\degree$.

Recent analyses~\cite{AtmRecent} of SuperKamiokande data
that make use of MonteCarlo predictions for the angular and energy
distributions of the atmospheric neutrinos
get more stringent constraints on the neutrino oscillation parameters.
Our fit uses only those data
--- the ratio of upward and downward multi-GeV neutrinos
(the ones in bins 1 and 5 of the angular distribution in~\cite{SK-atm}) ---
that do not depend
on the spectrum of the atmospheric neutrinos nor on the precise value of $\Delta m^2$,
assuming a full averaged oscillation in between.
Since statistics gives presently the dominant error,
we obtain weaker constraints than in~\cite{AtmRecent}.
If we knew that the neutrino mass difference relevant for
atmospheric neutrinos were close to the center of the presently allowed region,
we could add to the data to be fitted the intermediate bins 2 and 4 of~\cite{SK-atm}
(the bins that contain `oblique' neutrinos).
We cannot use in any case the multi-GeV data in the intermediate bin 3,
that contains `horizontal' neutrinos.
Having doubled the statistics, we would find the more stringent contours
shown in fig.~\ref{fig:Atm}${\rm a}',{\rm b1}',{\rm b2}'$.
We remind the reader that our fit does not include matter effects~\cite{MSW}.

\section{Atmospheric and Solar Neutrinos: Non-Minimal Schemes}\reseteq\label{AtmNonStandard}
In this section, we study atmospheric neutrinos when two conditions apply.
\begin{itemize}
\item The smallest mass splitting is large enough to affect
atmospheric neutrino oscillations:  $\Delta m^2_{12} > 3 \times
10^{-4}\eV^2$. For solar neutrinos, this implies that there is a serious
flaw either in at least one measurement technique, or in the solar models.
\item The mass splittings are hierarchical $\Delta m^2_{23} \gg
\Delta m^2_{12}$. This is a simplification, which we relax at the end
of the section. It includes the interesting possibility that  $\Delta m^2_{23}$
is large enough to induce the apparent oscillations reported by the
LSND collaboration~\cite{LSND}, while  $\Delta m^2_{12}$ effects are causing both
solar and atmospheric oscillations.
\end{itemize}
Using (\ref{eq:V}), the $\nu_e$ survival probability is
\begin{equation}
P_{ee} = 1 - c_{13}^4 \sin^2 2 \theta_{12} \; S_{12}
           - s_{12}^2 \sin^2 2 \theta_{13} \; S_{23}
           - c_{12}^2 \sin^2 2 \theta_{13} \; S_{31}
\label{eq:Pee}
\end{equation}
where $S_{ij} = \sin^2(\Delta m_{ij}^2t /4E)$.
The above two conditions imply that  $\Delta m^2_{23} > 2 \times 10^{-3}\eV^2$, so that, for the CHOOZ experiment, (\ref{eq:Pee}) should be
used with $S_{23} = S_{31} = 0.5$. The CHOOZ limit, $P_{ee} > 0.9$,
then gives $\theta_{13} < 0.23$.
If $\Delta m^2_{12}$ were also greater than $2 \times 10^{-3}\eV^2$,
then for the
CHOOZ experiment one also has $S_{12}=0.5$, so that $\theta_{12} < 0.23$.
However, in this case the survival probability for solar neutrinos is
the same as for the anti-neutrinos at CHOOZ: $P_{ee}> 0.9$. Hence,
given our two conditions, the observed solar neutrino fluxes require
$\Delta m^2_{12} < 2 \times 10^{-3}\eV^2$.

It is frequently stated that the three known neutrinos cannot explain
the LSND, atmospheric and solar neutrino anomalies, as this would
require three $\Delta m^2$ with different orders of magnitudes.
However, this argument no longer applies in the case that either a solar
neutrino measurement technique or solar models are incorrect, when a
single $\Delta m^2$ could give both atmospheric and solar anomalies.
Hence, we consider first the case that $\Delta m^2_{23}$ is large enough to
explain the observations of LSND.
The oscillation probabilities induced by  $\Delta m^2_{23} $
are given by (\ref{eq:P}). From the limit on $P_{ee}$ from the
Bugey reactor, one then concludes
\begin{eqnsystem}{sys:NonMinAtm}
\Delta m^2_{23} &>& 0.2\eV^2,
\label{eq:Deltam23}\\
\riga{and}\\[-2mm]
\theta_{13} &<& 0.1\label{eq:theta13}
\end{eqnsystem}
which is significantly stronger than the CHOOZ limit.
A second possibility, $\theta_{13}$ close to $90\degree$ does not allow
any significant oscillations of $\nu_{e}$ and is thus not acceptable
to explain the solar neutrino anomaly at a relatively large frequency.
For atmospheric
neutrinos, both upward going and downward going, one may then use
oscillation probabilities with $\theta_{13} = 0$\footnote{In which
case the $P_{ij}$ are independent of $\phi$.},
and with $S_{23}$ and $S_{13}$ both averaged to 0.5:
\begin{eqnsystem}{sys:PatmNonStandard}
P_{e \mu}    & = & c_{23}^2 \sin^2 2 \theta_{12}   \; S_{12}   \\
P_{e \tau}   & = & s_{23}^2 \sin^2 2 \theta_{12}   \; S_{12}    \\ 
P_{\mu \tau} & = & \textstyle - {1 \over 4}  \sin^2 2 \theta_{23} \sin^2 2
\theta_{12} \; S_{12} + {1 \over 2} \sin^2 2 \theta_{23} \label{eq:Pave}\\
\riga{or equivalently, from unitarity}\\[-4mm]
P_{e e}       & = &1- \sin^2 2 \theta_{12}   \; S_{12}  \\ 
P_{\mu \mu} & = &\textstyle 1 - {1 \over 2} \sin^2 2 \theta_{23}
- c_{23}^4 \sin^2 2 \theta_{12} \; S_{12}  \\
P_{\tau \tau}   & = & \textstyle 1 - {1 \over 2} \sin^2 2 \theta_{23}
 - s_{23}^4 \sin^2 2 \theta_{12} \; S_{12}.
\label{eq:Pdiagave}
\end{eqnsystem}
Since in these formul\ae{} $S_{13}=S_{23}$,
$\theta_{12}$ enters only via $\sin^2 2
\theta_{12}$ so that, without loss of generality,  we may reduce the range of $\theta_{12}$ to
$0 \le \theta_{12} < \pi/4$.
We again study the up-down ratios (\ref{eq:ratio}), as they have small
systematic uncertainties. We calculate them approximately, using
(\ref{eq:ratio}) with $f^e_{eCC}=f^\mu_{\mu CC}=f^{PC}_{\mu CC} =1$
and all other $f$-factors equal to zero.
A fraction, $P^{(0)}_{\mu \mu} = 1 -\sin^2 2 \theta_{23}/2$, of the
downward going $\nu_\mu$
oscillate to $\nu_\tau$ before detection, so the up-down ratios are
given by
\begin{equation}
\rho_e \approx P_{ee} + r P_{e \mu}
\label{eq:rhoe}
\end{equation}
and
\begin{equation}
\rho_\mu \approx {P_{\mu \mu} + P_{e \mu}/r \over
P^{(0)}_{\mu \mu}}.
\label{eq:rhomu}
\end{equation}
Hence we find
\begin{equation}
(\rho_\mu - 1) \approx -  {1 \over r} 
{c_{23}^2 \over 1 - {1 \over 2}
\sin^2 2 \theta_{23}} \cdot  (\rho_e - 1).
\label{eq:rhoratio}
\end{equation}
For the multi-GeV data, where the angular correlation is best, $r$ is
large, and (\ref{eq:rhoratio}) implies that $|\rho_\mu - 1| < (1/3)
|\rho_e - 1|$, in strong disagreement with data of (\ref{eq:ratioexp}).
The same
inequality holds if $\rho_\mu$ is replaced by $\rho_{PC}$, when the
disagreement with data is even stronger.\footnote{Even ignoring $\rho_e$,
we find $\rho_{PC,\mu} > 0.61$.}
{\em With oscillations of the three known neutrinos, the LSND
observation conflicts with the atmospheric and solar neutrino
anomalies even using a model independent analysis of the solar
neutrino fluxes or allowing for a systematic error in one of the
solar neutrino experiments\footnote{For $\theta_{23}=0$, this
corresponds to purely $\nu_\mu \rightarrow \nu_e$ oscillations, which is
therefore excluded as an explanation of the atmospheric neutrino
measurements.}.}

\smallskip

Does the atmospheric neutrino data allow other values of $\Delta
m^2_{23} \gg \Delta m^2_{12}$? The limit from the Bugey reactor,
(\ref{eq:theta13}),
applies for all $\Delta m^2_{23} > 0.06\eV^2$, and the up-down ratio
relation, (\ref{eq:rhoratio}), applies for all $\Delta m^2_{23} >
0.1\eV^2$. Hence,  $\Delta m^2_{23} > 0.1\eV^2$ is excluded. For
$\Delta m^2_{23} < 0.1\eV^2$, the downward going $\nu_\mu$ have not
oscillated to $\nu_\tau$ when they reach the Super-Kamiokande
detector, so that (\ref{eq:rhomu}) is replaced by
\begin{equation}
\rho_\mu \approx P_{\mu \mu} + {1 \over r} P_{e \mu}
= 1 - {1 \over 2} \sin^2 2 \theta_{23} - { c_{23}^2 \over r} (\rho_e -1).
\label{eq:rhomu2}
\end{equation}
Consistency with the data,~(\ref{eq:ratioexp}), is now possible, and
requires large $\theta_{23}$. As $\Delta m^2_{23}$ drops below $0.06\eV^2$,
the limit from the Bugey reactor on $\theta_{13}$ is
progressively weakened, so that $\theta_{13}$ terms must be kept in $P_{ij}$.
Furthermore, as $\Delta m^2_{23}$ drops below $0.01\eV^2$, our
hierarchy condition is no longer satisfied, so that $P_{ij}$ depend
also on $\theta_{12}$. For these cases we have performed a $\chi$ squared fit
of the three up-down ratios~(\ref{eq:ratioexp}) to $\theta_{23},
\theta_{13}$ and $\theta_{12}$, for various values of the mass
splittings, and have found acceptable regions of parameter space.
Results are shown in figure~\ref{fig:Atm}b for the case that all
$S_{ij}=0$ for downward going neutrinos, while all $S_{ij} = 0.5$ for
upward going neutrinos and $\phi=0$ (no CP violation).
An equivalent fit would be obtained for $\phi=\pi$ and $\theta_{23}\to\pi-\theta_{23}$.
The (relatively small) asymmetry of fig.s~\ref{fig:Atm}b under
$\theta_{23}\to\pi-\theta_{23}$ shows the dependence on $\phi$ of
the SuperKamiokande data considered here.

A comparison of figure~\ref{fig:Atm}b with figures~\ref{fig:Sun}
and \ref{fig:SunExpLess} shows under what conditions this large
$\Delta m^2_{12}$ scheme gives consistency. If all solar measurement
techniques are correct, then, from figure~\ref{fig:Sun},
$\theta_{13}$ is small and $\theta_{12} = 10\degree \div 20\degree$.
Figure~\ref{fig:Atm}b then shows that $\theta_{23}$ is centred on
$45\degree\pm25\degree$, the range around $\theta_{23}=135\degree$
being equivalent for any $\phi$ since $\theta_{13}$ is small.
Figure~\ref{fig:SunExpLess}
shows that solar model inspired fits to data from two solar techniques
at large $\Delta m^2_{12}$ allow larger ranges of $\theta_{12}$ and
$\theta_{13}$, and these become correlated with $\theta_{23}$ via
figure~\ref{fig:Atm}b.

\section{Large $\nu_\mu \rightarrow \nu_\tau$ Mixing For Atmospheric Neutrinos}
\label{matrices}\reseteq

The pattern of masses and mixings suggested by the previous considerations
show peculiar features, especially if both the atmospheric and solar neutrino anomalies
are accounted for in the minimal scheme of section~\ref{AtmStandard}.
The mass differences are hierarchical.  However
a large mixing  ($\theta_{23}\approx 45\degree$) is required 
between the states with the {\em largest} mass difference.
The mixing angle $\theta_{12}$ between the
states with the {\em smallest} mass splitting may be large or small.
Finally, if 
$\Delta m^2_{\rm atm} \geq 2\cdot10^{-3}\eV^2$, i.e. in the CHOOZ range, 
the third mixing angle must be small, $\theta_{13} \leq 13 \degree$.
Therefore it looks interesting to see
which mass matrix could produce this pattern
and which flavour symmetries can justify it.

\subsection{$2 \times 2$ Matrix Forms}\label{2x2}
As stressed in the introduction, an important consequence of the data
on atmospheric neutrino fluxes is the need for large mixing
angles. Here we study four possible forms of the $2 \times 2$ Majorana
mass matrix for $\nu_\mu$ and $\nu_\tau$ which have a large mixing angle.
In subsection~\ref{3x3} we study whether these forms can be
incorporated in $3 \times 3$ mixing schemes which also give solar
neutrino oscillations, and whether $3 \times 3$
cases exist which cannot be reduced to a $2 \times 2$ form.
In section~\ref{models} we study whether these forms may be obtained from flavour
symmetries of abelian type.

In a basis with a diagonal charged lepton mass matrix,
the Majorana neutrino mass matrix is
\begin{equation}
m = {v^2 \over M} \pmatrix{C & B \cr B & A}.
\label{eq:2by2m}
\end{equation}
This is brought into real, diagonal form by the unitary matrix
\begin{equation}
V = R(\theta) \pmatrix{1 & 0 \cr 0 & e^{i \alpha}}
\label{eq:2by2v}
\end{equation}
where $\tan 2 \theta = 2B/(A-C)$, and the phase $\alpha$ does not
affect oscillations. The mass difference relevant for oscillations is
$\Delta m^2 = (A+C) \sqrt{(A-C)^2 + 4B^2}$.
The coefficient $v^2/M$ is motivated by the see-saw
mechanism, with $v$ the electroweak vacuum expectation value and $M$
the mass of a heavy right-handed neutrino.

\begin{table}[t]
 \renewcommand{\arraystretch}{1.5}
 \newcommand{\lw}[1]{\smash{\lower2.ex\hbox{#1}}}
 \begin{center}
  \begin{tabular}{|l|c|cll|} \hline 
&\Blue small entries \Black
&\Blue small parameters \Black
&\Blue order unity parameters\Black
&\Blue {$\Delta m^2 /( {v^2 \over M} )^2$\Black } \cr \hline
\Blue(1)~~Generic\Black & none & none & $A, B, C$ & $\approx 1$ \cr
\Blue(2)~~Determinant small\Black & none  & none  & $A, B, C={B^2/ A} +
 \eps$ & $\approx 1$ \cr
\Blue(3)~~One diagonal small\Black & one diagonal  & $C \approx \eps$  & $A, B$  &
 $\approx 1$ \cr
\Blue(4)~~Pseudo-Dirac\Black & both diagonal  & $A,C \approx \eps$  & $B$  &
 $\approx \eps$ \cr
\hline 
\end{tabular}
\end{center}
\caption{\em The four possible $2 \times 2$ matrix forms which give a large mixing angle.}
\label{tb:2by2}
\end{table}

There are four possible forms of this matrix which give $\theta \approx 1$,
and these are shown in Table 1. In
cases (1) and (2) the entries are all of order unity; in the generic
case they are unrelated, while in case (2) they are related in such a
way that the determinant is suppressed. We discuss how such a
suppression can occur naturally via the seesaw mechanism in the next
section. Case (3) has one of the diagonal entries suppressed, which,
however, does not follow from a simple symmetry
argument. For cases (1$\div$3), taking $\Delta m^2 = 10^{-3}\eV^2$,
one finds
\begin{equation}
M = (1 \div 3) \times 10^{15} \mbox{GeV},
\label{eq:mgut}
\end{equation}
close to the scale of gauge coupling unification in supersymmetric
theories.

Finally, case (4) has both diagonal entries small, making  $\nu_\mu$
and $\nu_\tau$ components of a pseudo-Dirac neutrino. This follows
from an approximate $L_\mu - L_\tau$ symmetry, and implies that
$\theta \simeq 45\degree$. This agrees well with data: combing
$\rho_\mu$ and $\rho_{PC}$ of (\ref{eq:ratioexp})  gives
$\theta =45\degree \pm 15\degree$.
Of the four possible cases with large mixing angle, it is only the pseudo-Dirac
neutrino which allows $\nu_{\mu,\tau}$ to be the astrophysical hot
dark matter, in which case one predicts $\theta = 45\degree$ to high
accuracy.

From the viewpoint of atmospheric neutrino oscillations alone, the
distinction between cases (1) and (2) is unimportant. Since case (3)
does not follow from simple symmetry arguments, one is left with two
main $2 \times 2$ mixing schemes: the generic and pseudo-Dirac cases.


\subsection{$3 \times 3$ Matrix Forms}\label{3x3}
There are many possibilities for $3 \times 3$ neutrino mixing giving
$P_{\mu \mu} \approx 0.5$, with oscillation primarily to
$\nu_\tau$. In general two independent frequencies and three Euler
angles are involved.

For the case that the oscillation is dominated by a single frequency,
the possibilities may be divided into two classes: ``$2 \times
2$-like'' and ``inherently $3 \times 3$.''
The $2 \times 2$-like cases are just the four discussed in subsection~\ref{2x2},
with $\theta_{12,13}$ small. Even though $\Delta m^2_{23}$ may not be
the largest $\Delta m^2$, it is the only one which causes substantial
depletion of $\nu_\mu$. More interesting are the inherently $3 \times
3$ cases, for which there is no $2 \times 2$ reduction.

Consider the case
\begin{equation}
m = { v^2 \over M} \pmatrix{0&B&A \cr B&0&0 \cr A&0& 0}
\label{eq:3by3}
\end{equation}
with $A,B \approx 1$. This is diagonalized by $V = R_{23}(\theta_{23})
R_{12}(\theta_{12}=45\degree)$ giving a Dirac state of $\nu_e$
married to $c_{23} \nu_\mu + s_{23} \nu_\tau$. The mass eigenvalues
are $(M,M,0)$, which, from the viewpoint of oscillations are
equivalent to $(0,0,M)$. Hence, one immediately sees that the
oscillation probabilities are given by (\ref{eq:P}) with $\theta_{13}=0$:
$P_{\mu \tau} = \sin^2 2 \theta_{23} S_{23}$ has the form of a $2
\times 2$ oscillation, even though the mass matrix has an inherently
$3 \times 3$ form. This arises because (\ref{eq:3by3}) is governed by
the symmetry $L_e - L_\mu - L_\tau$, which allows $\nu_\mu
\leftrightarrow \nu_\tau$, but prevents $\nu_e$ from oscillating.

We claim that\eq{3by3} is the only inherently $3 \times 3$ form for $\nu_\mu
\rightarrow \nu_\tau$ {\em at a single frequency},
as we now show.
An inherently $3 \times 3$ form must have large entries outside the $2\times2$ block 
in 23 subspace.
The three possibilities are 11, 12 and 13 (and their symmetric).
None of these entries work alone, even coupled to any structure in the 23 block:
either one gets two comparable frequencies or one does not get $\nu_\mu\to\nu_\tau$.
The same is true for $11+12$ or $11+13$, again possibly together with any 23-block.
Since $11+12+13$ leads to two comparable frequencies, the only case remaining is $12+13$,
with a relatively negligible 23 block, i.e.\ the $3\times3$ form in\eq{3by3}.
Basic to this conclusion is the assumption of no special relations among
the different neutrino matrix elements other than
the symmetry of the matrix itself (for alternatives see~\cite{altrimenti}).

\section{Models for both Solar and Atmospheric Neutrinos}\label{models}\reseteq
In this section we construct models for the minimal scheme for atmospheric
and solar neutrino oscillations, discussed in section~\ref{AtmStandard}. 
The mass pattern suggested by this scheme has
the hierarchy $ \Delta m^2_{\odot} \equiv \Delta m^2_{12} \ll
\Delta m^2_{\rm atm} \equiv \Delta m^2_{23}$.
We take the form of the lepton mass matrices to be determined by
flavour symmetries (FS) and assume that all small entries in these
matrices are governed by small flavour symmetry breaking (FSB) parameters.

The low energy effective mass matrix for the three light left-handed
neutrinos can be written as the sum of two matrices:
$\mnuLL = m_{\rm atm} + m_\odot$, where all non-zero entries of $m_{\rm atm}$ are
larger than all entries of $m_\odot$. The form of $m_{\rm atm}$ is
such that there is a large mass splitting: $\Delta m^2_{\rm atm} \approx
10^{-(2\div 3)} \eV^2$, and a vanishing $\Delta m^2$.
Furthermore, this matrix must give a large
depletion of $\nu_\mu$, and, as discussed in the last section, this
could occur if it has certain $2 \times 2$-like or inherently
$3\times3$ forms. Of the two $2 \times 2$-like forms shown in Table 1,
only case (2) is acceptable: in cases (1) and (3) the two independent
$\Delta m^2$ are comparable, while in case (4) the second independent
$\Delta m^2$ is larger than $\Delta m^2_{\rm atm}$. Hence, we arrive at
the possibility\footnote{Ans\"atze 
of this type for the neutrino mass matrix, up to small corrections,
to describe atmospheric and solar neutrinos are contained in ref.s~\cite{2x2}.}:
\begin{equation}
m_{\rm atm}^{2 \times 2} = {v^2 \over M} \pmatrix{0 & 0 & 0 \cr
                                0 & C & B \cr 0 & B & A}
\label{eq:2by2form}
\end{equation}
with $A,B \approx 1$ and $C = B^2/A$.
A reason for the vanishing sub-determinant will be given shortly.

In the previous section we have proved that there is a unique form for $m_{\rm atm}$
which is inherently $3 \times 3$:
\begin{equation}
m_{\rm atm}^{3 \times 3} = {v^2 \over M} \pmatrix{0 & B & A \cr
                                B & 0 & 0 \cr A & 0 & 0}
\label{eq:3by3form}
\end{equation}
with $A,B \approx 1$.

The oscillation
angles in the leptonic mixing matrix, $V$, have contributions from
diagonalization of both the neutrino mass matrix, $\theta_{ij}^\nu$,
and the charged lepton mass matrix, $\theta_{ij}^e$:
$V(\theta_{ij}) = V^{e \dagger} (\theta^e_{ij}) V^\nu(\theta_{ij}^\nu)$.
This requires discussing also the charged lepton mass matrix.
It is not easy to construct an exhaustive list of the possible
symmetries and their breaking parameters. This is partly because there
are both discrete and continuous symmetries with many choices for
breaking parameters; but is mainly because of a subtlety of the seesaw
mechanism. Let $\mnuRR$ and $\mnuLR$ be the most general Majorana and
Dirac mass matrices of the seesaw mechanism allowed by some
approximate symmetry. On forming the mass matrix for the light states,
$\mnuLL = \mnuLR m^{-1}_{RR} \mnuLR^T$, one discovers that $\mnuLL$
need not be the most general matrix allowed by the approximate
symmetry. This means that one cannot construct an exhaustive list by
only studying the symmetry properties of $\mnuLL$ --- it is necessary
to study the full theory containing the right-handed states.

A casual glance at (\ref{eq:2by2form}) and (\ref{eq:3by3form}) shows
that the flavor symmetry we seek, from the viewpoint of $\Delta L = 2$
operators,
does not distinguish $l_\mu$ from $l_\tau$, but does distinguish these
from $l_e$. There are many combinations of the three lepton numbers
$L_a$, and their subgroups, acting on $l_a$, which have this property. As
representative of this group, we choose the combination $L_e-L_\mu -
L_\tau$. We find it remarkable that this symmetry group can yield both
(\ref{eq:2by2form}) and (\ref{eq:3by3form}), depending on how it is
realized.

\subsection{$L_e - L_\mu - L_\tau$ realized in the Low Energy Effective Theory}
In the effective theory at the weak scale, we impose an approximate
$L_e - L_\mu - L_\tau$ symmetry, which acts on the weak doublets, $l_{e,
\mu, \tau}$, and is broken by small FSB parameters, $\eps$ and
$\eps'$ of charge +2 and -2, respectively, giving a neutrino mass matrix:
\begin{equation}
\mnuLL = {v^2 \over M}
\pmatrix{\eps' & 1 & 1 \cr
            1 & \eps &\eps \cr
            1 & \eps &\eps}
\label{eq:emutau2}
\end{equation}
Hereafter, the various entries of the matrices only indicate the corresponding order
of magnitude, allowing for an independent parameter for each entry.
This texture gives
\begin{eqnsystem}{sys:modelL}
m_1\approx m_2 \approx\frac{v^2}{M}\phantom{\eps}&&\Delta m^2_{12}\approx \frac{v^4}{M^2}(\eps+\eps')\\
m_3\approx \frac{v^2}{M}\eps && \Delta m^2_{23} \approx \frac{v^4}{M^2}\label{eq:dm23modL}\\
\riga{and}\\
&\makebox[0cm]{$
\theta_{23}^\nu \approx 1 \qquad
\theta_{13}^\nu \approx \eps \qquad
\theta_{12}^\nu = 45\degree.$}&
\label{eq:thetanu}
\end{eqnsystem}
While the texture gives only the order of magnitude of $\theta_{23}^\nu$,
it precisely predicts $\theta_{12}^\nu$ to be close to $45\degree$. If
the FSB parameters $\eps$ and $\eps'$ are taken to be
extremely small, this becomes an excellent candidate for the case of
``just so'' solar neutrino oscillations, with the prediction that
$\theta_{12} = 45\degree$. However, from figure~\ref{fig:Sun} it
follows that this model cannot give matter neutrino oscillations in
the sun, which requires $\sin 2 \theta_{12} \le 0.9$.
There are several contributions to the deviation of $\sin 2
\theta_{12}$ from unity, but they are all too small to reconcile
the discrepancy. A hierarchy in
$\Delta m^2$ requires $\eps, \eps' < 0.1$, and since
$\sin^2 2\theta_{12}^\nu \simeq 1 - (\eps - \eps')^2/8$, the deviation
of $\sin 2 \theta_{12}^\nu$ from 1 is negligible. After
performing the $\theta_{12}^\nu$ rotation, there are small
$\Ord(\eps)$ rotations in the 13 and 23 planes necessary to
fully diagonalize $\mnuLL$; these are too small to affect our
conclusions. The last hope is that there could be a significant
contribution to $\theta_{12}$ from diagonalization of the charged
lepton mass matrix. 
As mentioned above, the diagonalization of the charged 
lepton mass matrix has to be discussed anyhow.

Consistently with the symmetry structure of\eq{emutau2}, the most general form for the charged
lepton mass matrix, with a structure governed by abelian symmetries is
\begin{equation}
 m_E = \lambda v \pmatrix{\xi'\phantom{\eps} & \xi \eps' & \eps' \cr
                  \xi' \eps & \xi\phantom{\eps'} & 1 \cr
                  \xi' \eps & \xi\phantom{\eps'} & 1}
\label{eq:mE}
\end{equation}
when left (right) handed leptons are contracted to the left (right),
$\bar{e}_L m_E e_R$.
$(1,\xi,\xi')$ are the relative FSB parameters of $(\tau_R,\mu_R,e_R)$ with respect to
some other approximate FS, needed to describe the charged lepton mass hierarchies,
and $\lambda$ is the absolute FSB parameter of $\bar{\tau}_R\tau_L$.
Here we ignore the fact that non-abelian symmetries could modify this form, for example
by requiring some entries to vanish.

Diagonalization of\eq{mE} leads to
$$\theta^e_{23}\approx 1,\qquad
\theta^e_{13}\approx\eps',\qquad
\theta^e_{12}\approx\eps'$$
Therefore, altogether
\begin{equation}
\theta_{23}\approx 1,\qquad
\theta_{12}\approx\eps+\eps',\qquad
\theta_{12}\approx45\degree.
\end{equation}
Since $\sin^2 2\theta_{12}$ remains corrected only by quadratic terms in $\eps$ and/or $\eps'$,
we conclude that  $L_e - L_\mu - L_\tau$, realized as an approximate
symmetry of the low energy effective theory, can explain both
atmospheric and solar neutrino fluxes with a hierarchy of $\Delta
m^2$, most likely {\em only} for the case of ``just so'' vacuum solar
oscillations, in which case the scale of new physics, $M$, is close to
the gauge unification scale, and the FSB parameters are extremely
small: $\eps, \eps' \approx 10^{-7}$.
This result also applies when {\em any}
approximate FS of the low energy effective theory yields\eq{emutau2}.
In view of~(\ref{sys:modelL}), with $\Delta m^2_{23}\approx\Delta m^2_{\rm atm}\approx 10^{-(2\div3)}\eV^2$,
notice that all three neutrinos are cosmologically irrelevant.
Furthermore, the smallness of the 11 entry of\eq{emutau2}
makes the search for neutrino-less $2\beta$-decay uninteresting.

Comparing the
$\theta_{13}$ plots of figures~\ref{fig:Sun} and~\ref{fig:SunExpLess},
one finds that, with one experiment
excluded, the case of $\theta_{12}= 45\degree$ becomes allowed for a large
range of $\Delta m^2_{12}$, giving another application for this inherently
$3 \times 3$ form of the mass matrix.

\subsection{$L_e - L_\mu - L_\tau$ realized via the Seesaw Mechanism}
The seesaw mechanism~\cite{seesaw} allows a simple origin for the vanishing of the
$2 \times 2$ sub-determinant of (\ref{eq:2by2form}). Consider a single
right-handed neutrino, $N$, with Majorana mass $M$ and Dirac mass term
$v N(\cos \theta \nu_\tau + \sin \theta \nu_\mu)$, where $\theta
\approx 1$. Integrating out this single heavy state produces a single
non-zero eigenvalue in $\mnuLL$ --- giving (\ref{eq:2by2form}) with
$A=\cos^2 \theta, B = \cos \theta \sin \theta$ and $C= \sin^2 \theta$,
so that $AC=B^2$.

How could this carry over to a theory with three right-handed
neutrinos, $N_a$? As long as one of them, $N$ with the above mass
terms, is much lighter than the others, then it will give the dominant
contribution to $\mnuLL$, which will have (\ref{eq:2by2form}) as its
leading term. Clearly the key is that there be one right-handed
neutrino which is lighter than the others, and couples comparably to
$\nu_\mu$ and $\nu_\tau$.

This can be realized using $L_e - L_\mu - L_\tau$, with two
small FSB parameters $\eps~(+2)$ and $\eps'~(-2)$.
The right-handed neutrino mass matrix is
\begin{equation}\label{eq:RR}
\mnuRR = M \pmatrix{\eps' & 1 & 1 \cr
                        1 & \eps &\eps \cr
                        1 & \eps & \eps}
\end{equation}
and the Dirac mass matrices of neutrinos and charged leptons are
\begin{equation}\label{eq:RL}
m_{LR} = \lambda'v \pmatrix{\eta' & \eps\eta' & \eps\eta' \cr
                    \eps'\eta & \eta & \eta \cr
                    \eps' & 1 & 1}
\qquad\hbox{and}\qquad
m_E =  \lambda v \pmatrix{\xi'\eta' & \eps\xi\eta' & \eps\eta' \cr
                        \xi'\eps'\eta & \xi\eta &\eta \cr
                        \xi'\eps' & \xi & 1} 
\end{equation}
where, in analogy with\eq{mE}, we have introduced FSB parameters consistent with\eq{RR}.

For ease of exposition, let us first consider the case where all the $\eta$ and $\xi$ factors
are set equal to unity.
The crucial point is that
there is a massless right-handed neutrino in the limit $\eps\to0$.
Hence, taking $\eps$ small, and doing a rotation
in the 23 plane we have $2 \times 2$ sub-matrices
\begin{equation}
\mnuRR^{-1} = {1 \over M} \pmatrix{0 & 0 \cr 0 & { 1/\eps}},\qquad
m_{LR} =  \lambda'v \pmatrix{1 & 1 \cr1 & 1}
\label{eq:2by2RRinv}
\end{equation}
giving
\begin{equation}
\mnuLL = { (\lambda'v)^2 \over M} \pmatrix{ 1/\eps &1/\eps \cr1/\eps & 1/\eps}
\label{eq:2by2LL}
\end{equation}
where det $\mnuLL = 0$ at this order. In a theory with right-handed neutrinos,
$L_e - L_\mu - L_\tau$ leads to (\ref{eq:2by2form}).

Extending the analysis to $3 \times 3$ matrices is
straightforward. The inverse of $\mnuRR$
\begin{equation}
\mnuRR^{-1} = {1 \over M} \pmatrix{\eps &1 & 1 \cr
                        1 & \eps' & \eps' \cr
                        1 & \eps' & { 1 \over \eps}}
\label{eq:3by3RRinv}
\end{equation}
shows a pseudo-Dirac structure in the 12 subspace, which is preserved
in the light neutrino mass matrix:
\begin{equation}
\mnuLL = { (\lambda'v)^2 \over M} \pmatrix{\eps &1 & 1 \cr
                        1 & \eps' & 0 \cr
                        1 & 0 & { 1 \over \eps}}
\label{eq:LLfinal}
\end{equation}
where we have gone to a basis which diagonalizes the 23 subspace.
The parameters relevant for neutrino oscillation are
\begin{eqnsystem}{sys:modR}
&\makebox[0cm]{$\displaystyle
\theta_{23}^{e,\nu} \approx 1,\qquad
\theta_{13}^{e,\nu} \approx \eps,\qquad
\theta_{12}^\nu = 45\degree  ,\qquad
\theta_{12}^e \approx \eps$}&\label{eq:thetaR}\\
\riga{and}\\[-2mm]
&\makebox[0cm]{$\displaystyle
\Delta m^2_{23} \approx  { 1 \over \eps^2} { (\lambda'v)^4 \over M^2} ,\qquad
\Delta m^2_{12} \approx  (\eps + \eps') { (\lambda'v)^4 \over M^2}.$}\label{eq:dm2r}\\
\riga{
It is remarkable that $L_e - L_\mu - L_\tau$ has forced a pseudo-Dirac
structure in the 12 subspace as in its previous realization, again
giving $\theta_{12}$ near $45\degree$. The crucial difference is that
the pseudo-Dirac mass splitting is now a higher power in FSB than before}\\
&\makebox[0cm]{$\displaystyle
{\Delta m^2_{12} \over \Delta m^2_{23}} \approx \eps^2 (\eps+ \eps')
$}\label{eq:dmq12/23}
\end{eqnsystem}
rather than $\eps + \eps'$. This allows $\eps$ and $\eps'$
to be considerably larger than before, so $\sin 2 \theta_{12} < 0.8$
is now possible, allowing large angle MSW solar neutrino
oscillations. In this case the FSB parameters are not very small
$\eps, \eps' \approx 0.3 \div 0.5$, so that the mass of the
right-handed neutrinos is still quite close to the gauge coupling
unification scale.
Notice again the cosmological irrelevance of the neutrino masses.
For neutrino-less $2\beta$ decay searches $(\mnuLL)_{11} \approx 
\epsilon^3 (\Delta m^2_{23})^{1/2} \leq 10^{-2}\eV^2$.
Finally, $\eps'\approx 0.1$ and $\lambda'\approx 1$
can make $M$ exactly coincident with the unification scale.

\medskip

So far we have only produced models with large $\theta_{12}$.
However $L_e-L_\mu-L_\tau$ realized with the seesaw mechanism may also lead to small $\theta_{12}$,
using the FSB suppression factors in\eq{RL}.
Taking $\eta'\ll\eps'$ and $\eta\approx 1$, in an appropriate 23 basis gives
\begin{equation}
\mnuLL = {(\lambda' v)^2 \over M} \pmatrix{{\eta'}^2\eps &\eta' & \eta' \cr
                        \eta' & \eps' & 0 \cr
                        \eta' & 0 & { 1 \over \eps}}
\label{eq:mLLeta'}
\end{equation}
so that eq.s\eq{dm2r} and \eq{dmq12/23} remain valid but
\begin{equation}
\theta_{23}^{e,\nu} \approx 1,\qquad
\theta_{13}^{e,\nu} \approx \eta'\eps ,\qquad
\theta_{12}^e \approx \eps'
\end{equation}
and, most importantly
\begin{equation}
\theta_{12}^\nu\approx \eta'/\eps'
\end{equation}
which can make $\theta_{12}$ small.

\section{Conclusions}\label{Conclusions}

The solar and atmospheric neutrino anomalies, strengthened by the recent 
SuperKamiokande observations, can be interpreted as due to 
oscillations of the three known neutrinos. However there is still considerable 
allowed ranges of masses and mixing angles that can account for all these anomalies,
especially if a cautious attitude is taken with regard to the theoretical analysis
and/or the (difficult) experiments relevant to solar neutrinos. A further major element of
uncertainty is related to the relatively large range of values for the mass splitting 
that can account for the atmospheric neutrino anomaly.
We summarize our conclusions by considering a set of 
alternative hypotheses, related to these dominant uncertainties, with an eye to 
the experimental program that may lead to their resolution and eventually to 
the determination of the full set of neutrino oscillation parameters.

A critical value for $\Delta m^2_{23}$ is around $2 \cdot 10^{-3}\eV^2$
mainly because for larger values CHOOZ sets a
considerable constraint on the mixing pattern, but also 
because $ (1\div2) \cdot 10^{-3}\eV^2$ is frequently discussed as a typical sensitivity limit for various 
Long-Base-Line (LBL) neutrino experiments, like the one from KEK to SK, or the 
$\nu_\tau$ appearance experiments with a high energy beam from CERN to Gran Sasso 
or from Fermilab to Soudan.
On the other end, a value of $\Delta m^2_{12}<2\cdot10^{-4}\eV^2$,
as certainly required by a standard Solar Neutrino Analysis (SNA),
would make the corresponding oscillation frequency irrelevant to the SK experiment on atmospheric neutrinos.
On this basis we consider the following four possibilities,
none of which, we believe, can be firmly excluded at present.
They are graphically represented in fig.~\ref{fig:4casi}.

\begin{enumerate}

\item $\Delta m^2_{23} > 2 \cdot 10^{-3}\eV^2$ and $\Delta m^2_{12}<2\cdot10^{-4}\eV^2$.
Here a minimal scheme to describe both solar and
atmospheric neutrinos is required, as discussed in section~\ref{AtmStandard}, with 
$\Delta m^2_{23} \gg \Delta m^2_{{12}}$. Since $\Delta m^2_{12}$ is too small
to affect atmospheric and/or LBL experiments, in both cases eqs.~\ref{sys:PatmStandard} apply. The fit 
relevant to SK is given in fig.~\ref{fig:Atm}a, with the further constraint, from
CHOOZ, that $\theta_{13}$ is small, $\theta_{13} \leq 13 \degree$, and therefore
$\theta_{23} = 45\degree \pm 15\degree$. In turn $\theta_{12}$, together with
$\Delta m^2_{12}$, will have to be determined by solar neutrino
experiments. In this alternative, the neatest 
confirmation of the SK result would come from a $\nu_\tau$ appearance  LBL experiment.
At the same time, a dominant $\nu_\mu \rightarrow \nu_\tau$ oscillation should also lead 
to a signal in the KEK to SK $\nu_\mu$ disappearance experiment, with no appreciable $\nu_e$
appearance signal.

\item $\Delta m^2_{23} < 2 \cdot 10^{-3}\eV^2$ and $\Delta m^2_{12}<2\cdot10^{-4}\eV^2$.
The main difference with respect to the previous case
is that now $\theta_{13}$ is not constrained by CHOOZ,
and therefore, from fig.~\ref{fig:Atm}a, it
can be as large as $45 \degree$. This implies, from eqs.~\ref{sys:PatmStandard}, that the results
of both atmospheric and LBL experiments, with low enough $\nu_\mu$
energies to permit exploration of $\Delta m^2$ lower than $2 \cdot 10^{-3}\eV^2$,
may be affected by a significant $P_{\mu e} \not= 0$.
By the same token, an experiment
with low energy $\bar{\nu}_e$ extending the sensitivity of CHOOZ (e.g. Kam-LAND) may show a large
signal if $\theta_{13}$ is indeed large. In any event $P_{\mu \tau}$ will be significant. 
Finally, as in case 1., decoupling of solar and atmospheric neutrino oscillations 
implies that $\theta_{12}$ can only be determined by solar neutrino experiments,
with an analysis complicated by $\theta_{13}$ being potentially unconstrained
(see fig.s~\ref{fig:Sun}, upper row)

\begin{figure}[t]\begin{center}
\begin{picture}(10,4.5)
\putps(0,0)(0,0){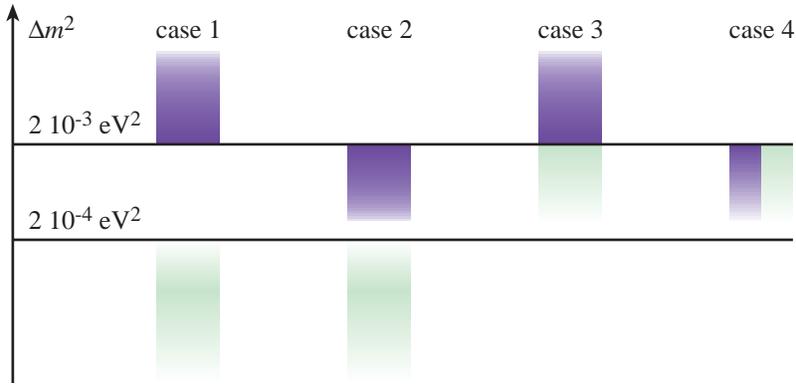}
\end{picture}
\caption[SP]{\em Different combinations of ranges for $\Delta m^2_{23}$ (dark gray)
and $\Delta m^2_{12}$ (light gray) discussed in the text.
\label{fig:4casi}}
\end{center}\end{figure}

\item $\Delta m^2_{23} > 2 \cdot 10^{-3}\eV^2$ and $\Delta m^2_{12}>2\cdot10^{-4}\eV^2$.
This case is possible only if SSM constraints are relaxed (fig.~\ref{fig:Sun}, lower row) and/or if one
of the experimental techniques for solar neutrinos is problematic (fig.~\ref{fig:SunExpLess}). 
However, as discussed in section~\ref{AtmNonStandard}, $\Delta m^2_{12}$ must be lower than $2 \cdot
10^{-3}\eV^2$, below the CHOOZ range. Since, on the other hand, $\Delta m^2_{\rm atm}
= \Delta m^2_{23}$ is in the CHOOZ range,
$\theta_{13}$ is small and eq.s~\ref{sys:PatmNonStandard} are relevant for  atmospheric and LBL
experiments. The fit of the present SK results gives 
$\theta_{23} = 45\degree \pm 25\degree$
(the range at $\theta_{23}\approx 135\degree$ being
equivalent since $\theta_{13}$ is small).
Therefore the main difference with respect 
to case 1.\ is the possibility of a $S_{12}$ contribution in eq.~(\ref{sys:PatmNonStandard}).
While $\nu_\tau$
appearance in LBL experiments must still give a positive signal, $P_{\mu e}$ could
significantly deviate from zero at low enough oscillation frequencies (relevant to
lower energy $\nu_\mu$ LBL experiments or to reactor experiments such as Kam-LAND).
The finding of such an effect, together with a positive $\nu_\tau$ appearance signal, would
prove, in the  three neutrino oscillation picture, the inadequacy of the NSA as it is
done now.

\item $\Delta m^2_{23}< 2 \cdot 10^{-3}\eV^2$ and $\Delta m^2_{12}>2\cdot10^{-4}\eV^2$.
This is the relatively less constrained case (and also the relatively less likely).
Here both neutrino squared mass differences are outside of the CHOOZ range, so that 
$\theta_{13}$ is unconstrained.
Appropriate values of the mixing angles can fit
the SuperKamiokande up/down ratios of atmospheric neutrinos,
as shown in fig.~\ref{fig:Atm}b.
In this case, the two comparable $\Delta m^2$ might lead to
sizeable CP-violating effects if all the three mixing angles are large.

\end{enumerate}
Measurements by SNO and Borexino will increase the number of independent
observational signals of the solar fluxes, $S_i$, from 3 to 5; so that,
from\eq{sigdep} with $\PhiCNO/\PhiBe = 0.22$, $\Delta m_{12}^2,
\theta_{12}, \theta_{13}, \PhiBe$ and $\PhiB$ can all be
determined. This will provide a crucial consistency check between the
experimental techniques and the solar models. If $\theta_{13}$ is found to
be large, $\Delta m^2_{23} < 2 \times 10^{-3}\eV^2$, giving a signal at
Kam-LAND, but making it harder for LBL experiments.

In the minimal scheme, with a hierarchy amongst the $\Delta m^2$, several
years of data from Super-Kamiokande will allow a fit to $\Delta m_{23}^2,
\theta_{23}$ and $\theta_{13}$. Combining  with fits to the solar flux
measurements, and to LBL and Kam-LAND experiments, could allow the
emergence of a consistent picture for the two oscillation frequencies and
the three leptonic mixing angles.

\smallskip

The variety of possibilities discussed above makes it uncertain which is the relevant neutrino
mass matrix and, a fortiori, which are the flavour symmetries that might be responsible
for it. Nevertheless, focusing on the minimal scheme for both solar and atmospheric 
neutrinos, the peculiar pattern of masses and mixings
 renders meaningful the search for an appropriate mass matrix.
As discussed in section~\ref{matrices} on general grounds,
two forms of mass matrices emerge as being able to describe the data,
eq.s\eq{2by2form} and\eq{3by3form}.
Since in the minimal scheme $\Delta m^2_{12} \ll 2 \cdot 10^{-3}\eV^2$,
these forms imply that neutrino masses will not give rise to
an observable neutrinoless double beta decay signal.
The combination $L_e - L_\mu - L_\tau$ of the individual lepton numbers may play a role 
in yielding both these forms. A common feature of the resulting solutions
is that the heaviest neutrino mass is determined by the oscillation
length of the atmospheric neutrinos, $(\Delta m^2_{\rm atm})^{1/2}$.
As such, the neutrino masses are irrelevant for present
cosmology. Again quite in general, an increasing separation
between the two $\Delta m^2$ requires the angle $\theta_{13}$ to become
increasingly small.

\paragraph{Acknowledgements}
We thank Zurab Berezhiani for a useful comment.
This work was supported
in part by the U.S. Department of Energy under Contracts DE-AC03-76SF00098,
in part by the National Science Foundation under grant PHY-95-14797,
in part by the TMR network under the EEC contract n.\ ERBFMRX-CT960090.

\setcounter{equation}{0}
\renewcommand{\theequation}{\thesection.\arabic{equation}}

\small~

\end{document}